\documentclass[12pt]{article}%
\usepackage{amsfonts}
\usepackage{graphicx}
\usepackage{amsmath}
\usepackage{amssymb}%
\pdfoutput=1
\makeatletter

\@addtoreset{equation}{section}
\makeatother
\newcommand{\beq}{\begin{equation}}
\newcommand{\eeq}{\end{equation}}
\renewcommand{\a}{\alpha}
\renewcommand{\b}{{{\beta}}}

\begin{document}
\baselineskip=18pt
\baselineskip 0.7cm

\begin{titlepage}
\setcounter{page}{0}
\renewcommand{\thefootnote}{\fnsymbol{footnote}}
\begin{flushright}
\end{flushright}
\begin{center}
{\LARGE \bf
Hodge Dualities on Supermanifolds\vskip 1cm
}
{\large
L. Castellani$^{~a,b,}$\footnote{leonardo.castellani@mfn.unipmn.it},
R. Catenacci$^{~a,c,}$\footnote{roberto.catenacci@mfn.unipmn.it},
and
P.A. Grassi$^{~a,b,}$\footnote{pietro.grassi@mfn.unipmn.it}\,.
\medskip
}
\vskip 0.5cm
{
\small\it
\centerline{$^{(a)}$ Dipartimento di Scienze e Innovazione Tecnologica, Universit\`a del Piemonte Orientale} }
\centerline{\it Viale T. Michel, 11, 15121 Alessandria, Italy}
\medskip
\centerline{$^{(b)}$ {\it
INFN, Sezione di Torino, via P. Giuria 1, 10125 Torino} }
\centerline{$^{(c)}$ {\it
Gruppo Nazionale di Fisica Matematica, INdAM, P.le Aldo Moro 5, 00185 Roma} }
\vskip  .3cm
\medskip
\end{center}
\smallskip
\centerline{{\bf Abstract}}
\medskip
\noindent
{
We discuss the cohomology of superforms and integral forms from
a new perspective based on a recently proposed Hodge dual operator.
We show how the superspace constraints
(a.k.a. rheonomic parametrisation) are translated from
the space of superforms $\Omega^{(p|0)}$ to the space of
integral forms $\Omega^{(p|m)}$ where $0 \leq p \leq n$, $n$ is the bosonic
dimension of the supermanifold and $m$ its fermionic dimension.
 We dwell on the relation between supermanifolds with non-trivial curvature
and Ramond-Ramond fields, for which the Laplace-Beltrami differential, constructed 
with our Hodge dual, is an essential ingredient. We discuss the definition of Picture Lowering
and Picture Raising Operators (acting on the space of superforms and on the space of integral forms)
and their relation with the cohomology. We construct non-abelian curvatures for gauge connections
in the space $\Omega^{(1|m)}$ and finally discuss Hodge dual fields within the present framework.

}
\vskip  .5cm
\noindent
{\today}
\end{titlepage}
\setcounter{page}{1}

\vfill
\eject
\tableofcontents

\vfill
\eject
\newpage\setcounter{footnote}{0} \newpage\setcounter{footnote}{0}

\section{Introduction}

Recently some issues regarding the structure and the properties of forms on supermanifolds
have been clarified. This has motivated the introduction of a new set of basic
forms (called integral forms) \cite{Voronov2,integ} which can be integrated on
supermanifolds and are useful for several applications in string theory and
quantum field theory
\cite{Berkovits:2004px,Berkovits:2009gi,Catenacci,Catenacci:2010cs,Witten:2012bg,Castellani:2014goa}. It has been shown also how the usual manipulations derived from the Cartan
calculus can be used. More recently, in \cite{Castellani:2015paa}, a
definition of Hodge dual has been proposed. It is based on the Fourier
transform for differential forms and is involutive. 
This new geometrical ingredient opens up the possibility of studying some
aspects of Hodge theory for differential forms on supermanifolds.

An important ingredient in the geometry of the supermanifolds is given by integral forms. 
As discussed in \cite{Witten:2012bg,Castellani:2014goa} they are essential
for a theory of integration of forms on supermanifolds. The wedge products of 
the differentials $d\theta$ ($\theta$ being the anticommuting coordinates) 
are commuting and therefore there is no canonical top form. To solve this problem, one introduces a
distribution-like quantity $\delta(d\theta)$ for which a complete Cartan calculus can be developed.
The distributions $\delta(d\theta)$ enter the definition of integral forms. 

The next step, tackled in \cite{Castellani:2015paa}, is the construction of a Hodge dual operator $\star$
for supermanifolds, that allows to apply well-known techniques such as Hodge theory
to study the cohomology classes of a given (super)manifold. In particular, for a compact bosonic
manifold $M$ endowed with a global metric $g$, one can introduce a nilpotent
differential operator  $d^\dagger = \star d \star$ (also called {\it codifferential}; $\star$ is the Hodge dual operator) and the Laplace-Beltrami differential $\Delta = d^\dagger\, d + d\, d^\dagger$. 
The latter is
used to compute harmonic forms ($\Delta \omega =0$) and the Hodge theorem states that for each de Rham cohomology class on $M$, there is a unique harmonic representative.
Obviously, due to invertibility of $\star$ for a given cohomology class of $d$ (de Rham cohomology) there exists a cohomology class of $d^\dagger$.

In the case of supermanifolds, the complex of pseudo-forms $\Omega^{(p|q)}$
 is filtered according to two numbers: $p$, the form number and $q$ the picture number.
 We have denoted by {\it superforms} those with vanishing picture, $\Omega^{(p|0)}$. They have no bound
 on the form degree. We have denoted by {\it integral forms} the complex $\Omega^{(p|m)}$, 
 where $m$ is the
 fermionic dimension of the supermanifold and $p \leq n$ (with $n$ the bosonic dimension). The $d$ operator
 can be conveniently extended; it increases the form number without touching the picture number. The Hodge dual operator maps
 a superform to an integral form (and therefore the picture number is changed). In particular, $\star$ maps $\Omega^{(p|0)}$ to $\Omega^{(n-p|m)}$. Therefore,
 $d^\dagger$ maps $\Omega^{(p|0)}$ to $\Omega^{(p-1|0)}$ and equivalently maps $\Omega^{(p|m)}$ to $\Omega^{(p-1|m)}$.
In terms of $d^\dagger$, we can finally define a Laplace-Beltrami differential.

In the present work, we study the codifferential $d^\dagger$ and the Laplace-Beltrami differential $\Delta$
on supermanifolds. Before doing that, we have to clarify
what type of cohomology we are describing. If we consider a flat supermanifold, applying the Poincar\'e lemma
(see for example \cite{Catenacci:2010cs}) it turns out that all closed superforms with positive form degree
are exact and therefore the cohomology coincides with the constant $(0|0)$-forms.
However, in the case of superforms, one can impose some external auxiliary conditions (known as superspace constraints or rheonomic conditions) which
make the cohomology non-trivial.  Let us consider an example: in three bosonic and two fermionic dimensions the gauge field belongs to a supermultiplet made out
of a vector field (with 3 d.o.f.'s, modulo the gauge transformations) and a spinor field (with 2 d.o.f.'s) . They can be cast into a vector superfield $A_a$ or into a spinor superfield $A_\a$ which are the components of a $1$-superform. Therefore, a $1$-superform contains too much freedom
for describing a single gauge supermultiplet. By
a clever choice of some constraints, one can preserve the covariance with respect to supersymmetry transformations,
reduce the number of independent superfields and restrict the physical field content to the one of a single gauge supermultiplet.

In addition, the computation of the cohomology with these external constraints
(which can be also be viewed as an {\it equivariant} cohomology) 
gives the irreducible representations of the superspin(\cite{Gates:1983nr}). 
Until now, the computation has been performed on the space of superforms $\Omega^{(p|0)}$, for which the form degree is not limited by the bosonic dimension of the supermanifold
(in the example, this is 3) and is in fact unbounded.  Potentially, at any given form degree we might have new cohomology classes and new multiplets. 
On the other side, we have seen that there is a new complex of forms $\Omega^{(p|m)}$ that, as we discuss in the following, is related by Hodge duality to the complex of superforms $\Omega^{(p|0)}$,  and 
therefore we should be able to relate the corresponding cohomology classes.
In addition, for this new complex there is a natural upper bound, but there is no lower bound. The complex of
integral forms $\Omega^{(p|m)}$ has maximum $p$ equal to the bosonic dimension and maximum 
$m$ equal to fermionic dimension of the supermanifold,
but, on the other hand, it contains also negative degree forms (by considering the derivatives of the Dirac delta forms). If there is a map
of cohomology classes between $\Omega^{(p|0)}$ and $\Omega^{(p|m)}$, it appears clear that the interesting cohomology classes must be contained
into the range of the non-trivial classes of both complexes.

In Sec. 2, some background material is given to set the stage and to fix the conventions used in the 
rest of the paper. In Sec. 2.2, we discuss the (super)Hodge dual operator for a generic metric on a given  supermanifold. In Sec. 2.3, the relation between Hodge theory, Laplace-Beltrami differential and the metric on supergroup manifolds is exploited. In Sec. 3 the complex of forms is studied and 
in Sec. 3.1, the horizontal differentials are constructed. On the other hand, in Sec. 3.2 the vertical 
differentials are introduced and discussed. In Sec. 4, the relation between superspace constraints for superforms and superspace constraints for integral forms is explored.  In Sec. 4.1 non-abelian 
field strengths are constructed. Finally Sec. 5  deals with dualities. 

\section{Background Material}
\subsection{$3d, N=1$}

We recall that in 3d N=1, the supermanifold $\mathcal{M}^{3|2}$ (homeomorphic
to $\mathbb{R}^{3|2}$) is described locally by the coordinates $(x^{a}%
,\theta^{\a})$, and in terms of these coordinates, we have the following two
differential operators
\begin{equation}
D_{\a}=\frac{\partial}{\partial \theta^{\a}}-\frac{1}{2}(\gamma^{a}\theta)_{\a}\partial_{a}%
\,,~~~~~~Q_{\a}=\frac{\partial}{\partial \theta^{\a}} +\frac{1}{2}(\gamma^{a}\theta)_{\a}\partial_{a}\,,~~~~~~
\label{susy3dA}%
\end{equation}
a.k.a. the superderivative and the supersymmetry generator, respectively. They have the
properties\begin{equation}
\{D_{\a},D_{\b} \}=-\gamma^a_{\a\b} \partial_a\,, ~~~~~~
\{Q_{\a},Q_{\b} \}=\gamma^a_{\a\b} \partial_a\,, ~~~~~~
\{D_{\a},Q_{\b} \}=0
\label{susy3dB}
\end{equation}
In 3d, we use real and symmetric Dirac matrices $\gamma^a_{\a\b}$. The conjugation matrix is
$\epsilon^{\a\b}$ and a bi-spinor is decomposed as follows: $R_{\a\b} = R \epsilon_{\a\b}  + R_a \gamma^a_{\a\b}$ , where
$R = - \frac12 \epsilon^{\a\b} R_{\a\b}$ and $R_a = {\rm tr}(\gamma_a R)$ are
a scalar and a vector, respectively.
In addition, it is easy to show that
$\gamma^{ab}_{\a\b} \equiv \frac12 [\gamma^a, \gamma^b]_{\a\b} = i \epsilon^{abc} \gamma_{c \a\b}$.

Given a $(0|0)$-form $\Phi^{(0|0)}$, we can compute its supersymmetry variation (viewed as a super translation)
as a Lie derivative $\mathcal{L}_{\epsilon}$ with $\epsilon=\epsilon
^{\a}Q_{\a}+\epsilon^{a}\partial_{a}$ ($\epsilon^{a}$ and  $\epsilon^{\a}$ are the infinitesimal
parameters of the translations in the $x$ and $\theta$ coordinates, respectively) and we have
\begin{equation}
\delta_{\epsilon}\Phi^{(0|0)}=\mathcal{L}_{\epsilon}\Phi^{(0|0)}%
=\iota_{\epsilon}d\Phi^{(0|0)}=\iota_{\epsilon}\Big(dx^{a}\partial_{a}%
\Phi^{(0|0)}+d\theta^{\a}\partial_{\a}\Phi^{(0|0)}\Big)=
\end{equation}%
\[
=(\epsilon^{a}+\frac{1}{2}\epsilon\gamma^{a}\theta)\partial_{a}\Phi
^{(0|0)}+\epsilon^{\a}\partial_{\a}\Phi^{(0|0)}=\epsilon^{a}\partial_{a}%
\Phi^{(0|0)}+\epsilon^{\a}Q_{\a}\Phi^{(0|0)}%
\]
In the same way, acting on $(p|q)$ forms, where $p$ is the form degree and $q$
is the picture number, we use the usual Cartan formula $\mathcal{L}_{\epsilon
}=\iota_{\epsilon}d+d\iota_{\epsilon}$.

To compute the differential of $\Phi^{(0|0)}$, we can use a set of
invariant $(1|0)$-forms
\begin{equation}
d \Phi^{(0|0)} = dx^{a} \partial_{a} \Phi^{(0|0)} + d\theta^{\alpha}%
\partial_{\alpha}\Phi^{(0|0)} =
\end{equation}
\[
=\Big(dx^{a} + \frac12 \theta\gamma^{a} d\theta\Big) \partial_{a} \Phi^{(0|0)}
+ d\theta^{\alpha}D_{\alpha}\Phi^{(0|0)} \equiv\Pi^{a} \partial_{a}
\Phi^{(0|0)} + \Pi^{\alpha}D_{\alpha}\Phi^{(0|0)}
\]
with the property of being invariant under supersymmetry transformations, namely $\delta_{\epsilon}\Pi^{m} = \delta_{\epsilon}\Pi^{\alpha}%
=0$. 

The particular top form represented by the expression
\begin{equation}
\label{top3d}\omega^{(3|2)} = \epsilon_{abc} \Pi^{a}\wedge\Pi^{b} \wedge
\Pi^{c} \wedge\epsilon^{\alpha\beta} \delta(d\theta^{\alpha}) \wedge
\delta(d\theta^{\beta})\,,
\end{equation}
has the properties:
\begin{equation}
\label{top3dB}d \omega^{(3|2)} = 0\,, ~~~~~ \mathcal{L}_{\epsilon}
\omega^{(3|2)} =0\,.
\end{equation}
It is important to point out the transformation properties of $\omega^{(3|2)}$ under a Lorentz transformation of $SO(2,1)$.
Considering $\Pi^a$, which transforms in the vector representation of $SO(2,1)$, the combination $\epsilon_{abc} \Pi^{a}\wedge\Pi^{b} \wedge \Pi^{c}$ is clearly invariant. On the other hand, $d\theta^\a$ transform under the spinorial
representation of $SO(2,1)$, say $\Lambda_\a^{~\b} = (\gamma^{ab})_\a^{~\b}  \Lambda_{ab}$ with $\Lambda_{ab} \in so(2,1)$, and thus
an expression like $\delta(d\theta^\a)$ is not covariant. Nonetheless, the combination $\epsilon^{\a\b} \delta(d\theta^\a) \delta(d\theta^\b) = 2 \delta(d\theta^1) \delta(d\theta^2)$ is invariant as can be proved using the properties of the
$\delta$-forms. In addition, $\omega^{(3|2)}$ has a bigger symmetry group: we can transform the variables $(\Pi^\a, d\theta^\a)$ under an element of the supergroup $SL(3|2)$. Note that $\omega^{(3|2)}$ is a section of the Berezinian bundle, the
equivalent for supermanifolds of the canonical bundle on bosonic manifolds.

In the case of supermanifolds, we can define the operator $\iota_X$ where $X$ are commuting or anticommuting vector fields. In the first case, the
operator $\iota_X$ is an anticommuting nilpotent operator acting on the sections of the exterior bundle. In the other case, if $X$ is anticommuting, then
$\iota_X$ is not nilpotent and is a commuting operator. For example 
we can choose $X = \partial_m$, namely the vector field
along the coordinate $x^a$, leading to $\iota_{\partial_a} \equiv \iota_a$ which is anticommuting 
$\{ \iota_a, \iota_b\} =0$. Alternatively, 
we can choose $X = \partial_\a$, the vector along the coordinate $\theta^\a$, and the corresponding 
$\iota_{\partial_\a} \equiv \iota_\a$
is a commuting operator $[\iota_\a, \iota_\b] =0$. It is convenient to represent $\iota_\a$ as a derivative with respect to $d\theta^\a$ since, loosely speaking, we are admitting any analytic function and distribution of those differentials (see also \cite{Witten:2012bg}).

In the following, we will use the definition of superforms, integral forms and pseudoforms. For that 
we recall here some of the main characteristics of them (more details are given in \cite{Castellani:2015paa}). We denote the spaces of forms by $\Omega^{(p|q)}$ where 
the index $p$ corresponds to form degree and $q$ is the picture number. A given 
form is expanded on a basis of $1$-form differentials $dx^a, d\theta^\a$ and 
distributional-like differentials $\delta(d\theta^\a)$, (notice that $\delta(dx^a) = dx^a$), as follows 
\begin{equation}
{\omega^{(p|q)}=\sum_{r=0}^{p}\omega_{\lbrack a_{1}\dots a_{r}](\alpha
_{r+1}\dots\alpha_{p})[\beta_{1}\dots\beta_{q}]}(x,\theta)dx^{a_{1}}\dots
{}dx^{a_{r}}d\theta^{\alpha_{r+1}}\dots{}d\theta^{\alpha_{p}}\delta
(d\theta^{\beta_{1}})\dots\delta(d\theta^{\beta_{q}})} \label{integralform}%
\end{equation}
with $\omega_{\lbrack a_{1}\dots a_{r}](\alpha_{r+1}\dots\alpha_{p})[\beta
_{1}\dots\beta_{q}]}(x,\theta)$ superfields.
The number $q$ counts the number of Dirac delta functions. In addition, we can also admit 
derivatives of delta functions which decrease the form degree. If $q =0$, then we call the space 
$\Omega^{(p|0)}$ the space of {\it superforms}, largely studied in the literature. For $q =2$ (namely 
the maximum number of fermionic coordinates in our example) the space $\Omega^{(p|2)}$ is called the 
space of {\it integral forms} (since they can be integrated on a supermanifold) and, finally, $\Omega^{(p|q)}$ for $0<q<2$ is the space of {\it pseudoforms}. 

\subsection{(Super)Hodge dual}

In this subsection we recall the construction of the Hodge dual for super and
integral forms. This construction was described in \cite{Castellani:2015paa};
we present here a modified (but similar) approach more explicit and more
suitable for physical applications.\footnote{In the paper
\cite{Castellani:2015paa} we started with the case of the Hodge dual for a
standard orthonormal basis in the appropriate exterior modules. This basis is
the one in which the supermetric is diagonal (not simply block diagonal).
Trasforming to a generic $\mathbb{Z}_{2}$- ordered basis we get the Hodge dual
for a generic block diagonal metric. This procedure and the one described in the
present paper give the same results.}

We consider a supermanifold $\mathcal{M}$ homeomorphic to $\mathbb{R}^{n|m}$
and we denote by $T$ the tangent bundle and by $T^{\ast}$ the cotangent bundle.

To simplify the notations we will denote by the same letter a bundle and the
$\mathbb{Z}_{2}-$graded modules of its sections. These modules are generated over the ring of superfunctions as follows
( $d$ is an odd derivation and $i=1...n\text{ ; }\alpha
=1...m$) :
\begin{align*}
&  T\text{ by the even vectors }\frac{\partial}{\partial x^{i}}\ \text{and the
odd vectors }\frac{\partial}{\partial\theta^{\alpha}}\\
&  T^{\ast}\text{ by the even forms }d\theta^{\alpha}\text{and the odd forms
}dx^{i}%
\end{align*}
If $\Pi$ is the parity reversal symbol $\left(  \Pi\mathbb{R}^{p|q}%
=\mathbb{R}^{q|p}\right)  $, we can consider the bundle $\Pi T$. The
$\mathbb{Z}_{2}-$graded module of its sections is generated by the even
vectors $b_{\alpha}$ and the odd vectors $\eta_{i}.$ Using the (super) wedge
products:%
\begin{align*}
dx^{i}dx^{j}  &  =-dx^{j}dx^{i}\,,\quad dx^{i}d\theta^{\alpha}=d\theta
^{\alpha}dx^{i\,},\quad d\theta^{\alpha}d\theta^{\beta}=d\theta^{\beta}%
d\theta^{\alpha},\\
\theta^{\alpha}dx^{i}  &  =-dx^{i\,}\theta^{\alpha},\quad\theta^{\alpha
}d\theta^{\beta}=d\theta^{\beta}\theta^{\alpha}\\
\eta_{i}\eta_{j}  &  =-\eta_{j}\eta_{i}\,,\quad\eta_{i}b_{\alpha}=b_{\alpha
}\eta_{i},\quad b_{\alpha}b_{\beta}=b_{\beta}b_{\alpha},\\
\theta^{\alpha}\eta_{i}  &  =-\eta_{i}\theta^{\alpha},\quad\theta^{\alpha
}b_{\beta}=b_{\beta}\theta^{\alpha}%
\end{align*}
we can construct the super exterior bundles $\wedge T^{\ast}$ and $\wedge\Pi
T$ and we can give to the $\mathbb{Z}_{2}-$graded modules of the sections of
these bundles the structure of $\mathbb{Z}_{2}-$graded algebras, denoted again
by $\wedge T^{\ast}$ and $\wedge\Pi T.$

We consider now the $\mathbb{Z}_{2}-$ graded tensor product (over the ring of
superfunctions) $T^{\ast}\otimes\Pi T\ $and the invariant even section
$\sigma$ given by:%
\begin{equation}
\sigma=dx^{i}\otimes\eta_{i}+d\theta^{\alpha}\otimes b_{\alpha}
\label{sezionesigma}%
\end{equation}

If we define $A=g\left(  \frac{\partial}{\partial x^{i}},\frac{\partial}{\partial
x^{j}}\right)  $ to be a (pseudo)riemannian metric$\ $and\textbf{ }$B=\gamma
(\frac{\partial}{\partial\theta^{\alpha}},\frac{\partial}{\partial
\theta^{\beta}})$ to be a symplectic form, the even matrix $\mathbb{G}=%
\begin{pmatrix}
A & 0\\
0 & B
\end{pmatrix}
$ is a supermetric in $\mathbb{R}^{n|m}$ (with obviously $m$ even). $A$ and
$B$ are, respectively, $n\times n$ and  $m\times m$ invertible matrices
with real entries and $\det A\neq0$, $\det B\neq0.$

In matrix notations, omitting (here and in the following) the tensor product
symbol, the section $\sigma$ can be written as:%
\[
\sigma=dxAA^{-1}\eta+d\theta BB^{-1}b=dxA\eta^{\prime}+d\theta Bb^{\prime
}=dZ\mathbb{G}W^{\prime}%
\]
where $\eta^{\prime}=A^{-1}\eta$ and $b^{\prime}=B^{-1}b$ are the covariant
forms corresponding to the vectors $\eta$ and $b$; $dZ=\left(  dx\text{
}d\theta\right)  $ and $W^{\prime}=%
\begin{pmatrix}
\eta^{\prime}\\
b^{\prime}%
\end{pmatrix}
.$

If $\omega(x,\theta,dx,d\theta)$ is a superform in $\Omega^{\left(
p|0\right)  },$ the section $\sigma$ can be used to generate an integral transform%

\[
\mathcal{T}(\omega)=\int_{\mathbb{R}^{m|n}}\omega(x,\theta,\eta^{\prime
},b^{\prime})e^{i\left(  dxA\eta^{\prime}+d\theta Bb^{\prime}\right)  }\left[
d^{n}\eta^{\prime}d^{m}b^{\prime}\right]
\]
Where $\omega(x,\theta,\eta^{\prime},b^{\prime})$ has polynomial dependence in
the variables $\theta,\eta^{\prime}$ and $b^{\prime}$ and $e^{i\sigma}%
\in\wedge T^{\ast}\otimes\wedge\Pi T$ is a power series defined recalling that
if $\mathcal{A}$ and $\mathcal{B}$ are two $\mathbb{Z}_{2}$-graded algebras
with products $\cdot_{\mathcal{A}}$and $\cdot_{\mathcal{B}}$, the
$\mathbb{Z}_{2}$-graded tensor product $\mathcal{A}\otimes\mathcal{B}$ is a
$\mathbb{Z}_{2}$-graded algebra with the product given by (for homogeneous
elements);%
\[
(a\otimes b)\cdot_{\mathcal{A}\otimes\mathcal{B}}(a^{\prime}\otimes b^{\prime
})=(-1)^{\left\vert a^{\prime}\right\vert \left\vert b\right\vert }%
a\cdot_{\mathcal{A}}a^{\prime}\otimes b\cdot_{\mathcal{B}}b^{\prime}%
\]
In our case the algebras under consideration are the super exterior algebras
and the products $\cdot$ are the super wedge products defined above. We have,
for example:
\begin{align*}
(dx^{i}\otimes\eta^{\prime j})\cdot(dx^{l}\otimes\eta^{\prime k})  &
=-dx^{i}dx^{l}\otimes\eta^{\prime j}\eta^{\prime k}\\
(dx^{l}\otimes\eta^{\prime k})\cdot(dx^{i}\otimes\eta^{\prime j})  &
=(dx^{i}\otimes\eta^{\prime j})\cdot(dx^{l}\otimes\eta^{\prime k})\\
(1\otimes\eta^{\prime j})\cdot(dx^{l}\otimes\eta^{\prime k})  &
=-dx^{l}\otimes\eta^{\prime j}\eta^{\prime k}\\
(dx^{i}\otimes1)\cdot(dx^{l}\otimes\eta^{\prime k})  &  =dx^{i}dx^{l}%
\otimes\eta^{\prime k}\\
(1\otimes\eta^{\prime j})\cdot(dx^{l}\otimes1)  &  =-dx^{l}\otimes\eta^{\prime
j}\\
(dx^{l}\otimes1)\cdot(1\otimes\eta^{\prime j})  &  =dx^{l}\otimes\eta^{\prime
j}%
\end{align*}

In the following we will omit the symbols $\cdot$ and $\otimes.$

We have then:%
\[
e^{i\sigma}=e^{idxA\eta^{\prime}}e^{id\theta Bb^{\prime}}=\sum_{k=0}^{n}%
\frac{1}{k!}\left(  idxA\eta^{\prime}\right)  ^{k}e^{id\theta Bb^{\prime}}%
\]

The integral over the odd $\eta^{\prime}$ variables is a Berezin integral and
the integral over the even $b^{\prime}$ variables is defined by formal rules,
for example:%
\begin{subequations}
\begin{align}
\int_{\mathbb{R}^{m}}e^{id\theta Bb^{\prime}}d^{m}b^{\prime}  &  =\frac
{1}{\det B}\,\delta^{m}(d\theta)\label{rappintegrale}\\
\int_{\mathbb{R}^{m}}b_{1}^{\prime}...b_{m}^{\prime}e^{id\theta Bb^{\prime}%
}d^{m}b^{\prime}  &  =(-i)^{m}\,\frac{1}{\left(  \det B\right)  ^{m+1}}\left(
\frac{d}{d\theta}\delta(d\theta)\right)  ^{m} \label{rappintegrale1}%
\end{align}
The products $\delta^{m}(d\theta)$ and $\left(  \frac{d}{d\theta}%
\delta(d\theta)\right)  ^{m}$ ($m$ here denotes the number of factors) are
wedge products ordered as in $d^{m}b.$ In other words this kind of integrals
depends on the choice of an oriented basis. For example, we obtain the crucial
anticommuting property of the delta forms (no sum on $\alpha,\beta$):
\end{subequations}
\begin{equation}
\delta(d\theta^{\alpha})\delta\left(  d\theta^{\beta}\right)  =\int
_{\mathbb{R}^{2}}e^{i(d\theta^{\alpha}b^{\prime\alpha}+d\theta^{\beta
}b^{\prime\beta})}db^{\prime\alpha}db^{\prime\beta}=-\int_{\mathbb{R}^{2}%
}e^{i(d\theta^{\alpha}b^{\prime\alpha}+d\theta^{\beta}b^{\prime\beta}%
)}db^{\prime\beta}db^{\prime\alpha}=-\,\delta(d\theta^{\beta})\delta\left(
d\theta^{\alpha}\right)
\end{equation}

As observed in \cite{Castellani:2015paa} one can obtain the usual Hodge dual
in $\mathbb{R}^{n}$\textbf{ }(for a metric given by the matrix $A$\textbf{)
}by means of\textbf{ }the transform $\mathcal{T}$ . For $\omega(x,dx)\in
\Omega^{k}(\mathbb{R}^{n})$ we have:
\begin{equation}
\star\omega=i^{\left(  k^{2}-n^{2}\right)  }\frac{\sqrt{\left\vert
g\right\vert }}{g}\mathcal{T}(\omega)=i^{\left(  k^{2}-n^{2}\right)  }%
\frac{\sqrt{\left\vert g\right\vert }}{g}\int_{\mathbb{R}^{0|n}}\omega
(x,\eta^{\prime})e^{idxA\eta^{\prime}}[d^{n}\eta^{\prime}]\label{duale1}%
\end{equation}
where $g=\mathrm{det}A$. 

For example, in $\mathbb{R}^{2}$\textbf{ }we can compute:%
\begin{equation}
e^{idxA\eta^{\prime}}=1+ig_{11}dx^{1}\eta^{\prime1}+ig_{21}dx^{2}\eta
^{\prime1}+ig_{12}dx^{1}\eta^{\prime2}+ig_{22}dx^{2}\eta^{\prime2}%
+gdx^{1}dx^{2}\eta^{\prime1}\eta^{\prime2}\label{fourier5}%
\end{equation}
and the definition (\ref{duale1}) gives the usual results:
\begin{align*}
\star1 &  =i^{\left(  0^{2}-2^{2}\right)  }\mathcal{T}(1)=\frac{\sqrt
{\left\vert g\right\vert }}{g}\int_{\mathbb{R}^{0|2}}e^{idxA\eta^{\prime}%
}[d^{2}\eta^{\prime}]=\sqrt{\left\vert g\right\vert }dx^{1}dx^{2}\\
\star dx^{1}dx^{2} &  =i^{\left(  2^{2}-2^{2}\right)  }\mathcal{T}%
(\eta^{\prime1}\eta^{\prime2})=\frac{\sqrt{\left\vert g\right\vert }}{g}%
\int_{\mathbb{R}^{0|2}}\eta^{\prime1}\eta^{\prime2}e^{idxA\eta}[d^{2}%
\eta^{\prime}]=\frac{\sqrt{\left\vert g\right\vert }}{g}\\
\star dx^{1} &  =i^{\left(  1^{2}-2^{2}\right)  }\mathcal{T}(\eta^{\prime
1})=i^{\left(  1^{2}-2^{2}\right)  }\frac{\sqrt{\left\vert g\right\vert }}%
{g}\int_{\mathbb{R}^{0|2}}\eta^{\prime1}e^{idxA\eta^{\prime}}[d^{2}%
\eta^{\prime}]=-g^{12}\sqrt{\left\vert g\right\vert }dx^{1}+g^{11}%
\sqrt{\left\vert g\right\vert }dx^{2}\\
\star dx^{2} &  =i^{\left(  1^{2}-2^{2}\right)  }\mathcal{T}(\eta^{\prime
2})=i^{\left(  1^{2}-2^{2}\right)  }\frac{\sqrt{\left\vert g\right\vert }}%
{g}\int_{\mathbb{R}^{0|2}}\eta^{\prime2}e^{idxA\eta^{\prime}}[d^{2}%
\eta^{\prime}]=-g^{22}\sqrt{\left\vert g\right\vert }dx^{1}+g^{21}%
\sqrt{\left\vert g\right\vert }dx^{2}%
\end{align*}
The factor $i^{\left(  k^{2}-n^{2}\right)  }$ can be obtained by computing the
transformation of the monomial form $dx^{1}dx^{2}...dx^{k}\ $in the simple
case $A=I.$

Noting that in the Berezin integral only the higher degree term in the $\eta$
variables is involved, and that the monomials $dx^{i}\eta^{i}$ are even
objects, we have:%
\begin{align*}
\mathcal{T}\left(  dx^{1}...dx^{k}\right)   &  =\int_{\mathbb{R}^{0|n}}%
\eta^{1}...\eta^{k}e^{idx\eta}[d^{n}\eta]=\\
&  =\int_{\mathbb{R}^{0|n}}\eta^{1}...\eta^{k}e^{i\left(  \sum_{i=1}^{k}%
dx^{i}\eta^{i}+\sum_{i=k+1}^{n}dx^{i}\eta^{i}\right)  }[d^{n}\eta]=\\
&  =\int_{\mathbb{R}^{0|n}}\eta^{1}...\eta^{k}e^{i\sum_{i=1}^{k}dx^{i}\eta
^{i}}e^{i\sum_{i=k+1}^{n}dx^{i}\eta^{i}}[d^{n}\eta]=\\
&  =\int_{\mathbb{R}^{0|n}}\eta^{1}...\eta^{k}e^{i\sum_{i=k+1}^{n}dx^{i}%
\eta^{i}}[d^{n}\eta]=\\
&  =\int_{\mathbb{R}^{0|n}}\frac{i^{n-k}}{\left(  n-k\right)  !}\eta
^{1}...\eta^{k}\left(  \sum_{i=k+1}^{n}dx^{i}\eta^{i}\right)  ^{n-k}[d^{n}%
\eta]
\end{align*}
Rearranging the monomials $dx^{i}\eta^{i}$ one obtains:
\[
\left(  \sum_{i=k+1}^{n}dx^{i}\eta^{i}\right)  ^{n-k}=\left(  n-k\right)
!\left(  dx^{k+1}\eta^{k+1})(dx^{k+2}\eta^{k+2})...(dx^{n}\eta^{n}\right)  =
\]%
\[
=\left(  n-k\right)  !(-1)^{\frac{1}{2}(n-k)(n-k-1)}\left(  dx^{k+1}%
dx^{k+2}...dx^{n})(\eta^{k+1}\eta^{k+2}...\eta^{n}\right)
\]
Finally we obtain:%
\[
\mathcal{T}\left(  dx^{1}...dx^{k}\right)  =
\]%
\[
=\int_{\mathbb{R}^{0|n}}\frac{i^{n-k}}{\left(  n-k\right)  !}\eta^{1}%
...\eta^{k}\left(  n-k\right)  !(-1)^{\frac{1}{2}(n-k)(n-k-1)}\left(
dx^{k+1}dx^{k+2}...dx^{n})(\eta^{k+1}\eta^{k+2}...\eta^{n}\right)  [d^{n}%
\eta]=
\]%
\[
=\int_{\mathbb{R}^{0|n}}i^{n-k}(-1)^{\frac{1}{2}(n-k)(n-k-1)}(-1)^{k(n-k)}%
\left(  dx^{k+1}dx^{k+2}...dx^{n})(\eta^{1}...\eta^{k})(\eta^{k+1}\eta
^{k+2}...\eta^{n}\right)  [d^{n}\eta]=
\]%
\[
=i^{\left(  n^{2}-k^{2}\right)  }(dx^{k+1}dx^{k+2}...dx^{n})
\]
The computation above gives immediately:%
\begin{equation}
i^{\left(  k^{2}-n^{2}\right)  }\mathcal{T}\left(  dx^{1}...dx^{k}\right)
=\star\left(  dx^{1}...dx^{k}\right)
\end{equation}
and%
\begin{equation}
\mathcal{T}^{2}\left(  \omega\right)  =i^{\left(  n^{2}-k^{2}\right)
}i^{\left(  k^{2}\right)  }\left(  \omega\right)  =i^{n^{2}}\left(
\omega\right)  \label{quadratodifourier}%
\end{equation}
yielding the usual relation:%
\begin{equation}
\star\star\omega=i^{(\left(  n-k)^{2}-n^{2}\right)  }i^{\left(  k^{2}%
-n^{2}\right)  }i^{n^{2}}(\omega)=(-1)^{k(k-n)}(\omega)
\end{equation}

We can generalize the Hodge dual to superforms of zero picture (note that the
spaces of superforms or of  integral forms are all
finite dimensional) where we have the two types of differentials, $d\theta$
and $dx.$ The integral transform must be performed on the differentials:%
\begin{equation}
\mathcal{T}(\omega)=\int_{\mathbb{R}^{m|n}}\omega(x,\theta,\eta^{\prime
},b^{\prime})e^{i\left(  dxA\eta^{\prime}+d\theta Bb^{\prime}\right)  }\left[
d^{n}\eta^{\prime}d^{m}b^{\prime}\right]  \label{trasformatapersuperforme}%
\end{equation}

A zero picture $p-$superform $\omega$ is a combination of a \textbf{finite
number} of monomial elements of the form:%
\begin{equation}
\rho_{\left(  r,l\right)  }\left(  x,\theta,dx,d\theta\right)  =f(x,\theta
)dx^{i_{1}}dx^{i_{2}}...dx^{i_{r}}\left(  d\theta^{1}\right)  ^{l_{1}}\left(
d\theta^{2}\right)  ^{l_{2}}...\left(  d\theta^{s}\right)  ^{l_{s}}%
\end{equation}
of total degree equal to $p=r+l_{1}+l_{2}+...+l_{s}.$ We denote by $l$ the sum
of the $l_{i}.$ We have also $r\leq n.$

The super Hodge dual on the monomials can be defined as:
\begin{equation}
\star\rho_{\left(  r,l\right)  }=\left(  i\right)  ^{r^{2}-n^{2}}\left(
i\right)  ^{\alpha\left(  l\right)  }\mathcal{T(}\rho_{\left(  r,l\right)
})=\left(  i\right)  ^{r^{2}-n^{2}}\left(  i\right)  ^{\alpha\left(  l\right)
}\frac{\sqrt{\left\vert S\mathrm{\det}\mathbb{G}\right\vert }}{S\mathrm{\det
}\mathbb{G}}\int_{\mathbb{R}^{m|n}}\rho_{\left(  r,l\right)  }(x,\theta
,\eta^{\prime},b^{\prime})e^{i\left(  dxA\eta^{\prime}+d\theta Bb^{\prime
}\right)  }[d^{n}\eta^{\prime}d^{m}b^{\prime}]\label{superhodge1}%
\end{equation}
We recall that:
\[
S\mathrm{\det}\mathbb{G}=\frac{\mathrm{det}A}{\mathrm{det}B}%
\]

The normalization coefficient is given by: $\alpha(l)=2pl-l^{2}-nl-l$ (with
$l=p-r$) if $n$ is even and $\alpha(l)=l$ if $n$ is odd. These coefficient was
computed in \cite{Castellani:2015paa}

The $\star$ operator on monomials can be extended by linearity to generic
forms in $\Omega^{(p|0)}:$%

\[
\star:\Omega^{(p|0)}\longrightarrow\Omega^{(n-p|m)}%
\]
Both spaces are \textbf{finite dimensional} and $\star$ is an
isomorphism\footnote{The normalization coefficients chosen in the definitions
of the duals of $\rho_{\left(  r,l\right)  }$ and $\rho_{\left(  r|j\right)
}$ lead to the usual duality on $\Omega^{\left(  p|0\right)  }:$%
\[
\star\star\rho_{\left(  r,p-r\right)  }=(-1)^{p(p-n)}\rho_{\left(
r,p-r\right)  }%
\]
}.

An important example in $\mathbb{R}^{n|m}$ is $1\in\Omega^{(0|0)}$:%
\[
\star1=\sqrt{\left\vert \frac{\mathrm{det}A}{\mathrm{det}B}\right\vert }%
d^{n}x\delta^{m}(d\theta)\in\Omega^{(n|m)}%
\]

In the case of $\Omega^{(p|m)},$ a $m-$ picture $p-$integral form $\omega$ is
a combination of a \textbf{finite number} of monomial elements of the form:%
\begin{equation}
\rho_{\left(  r|j\right)  }\left(  x,\theta,dx,d\theta\right)  =f(x,\theta
)dx^{i_{1}}dx^{i_{2}}...dx^{i_{r}}\delta^{\left(  j_{1}\right)  }\left(
d\theta^{1}\right)  \delta^{\left(  j_{2}\right)  }\left(  d\theta^{2}\right)
...\delta^{\left(  j_{m}\right)  }\left(  d\theta^{m}\right)
\end{equation}
where $p=r-\left(  j_{1}+j_{2}+...+j_{m}\right)  .$ We denote by $j$ the sum
of the $j_{i}.$ We have also $r\leq n$.

The Hodge dual is:%
\begin{equation}
\star\rho_{\left(  r|j\right)  }=\left(  i\right)  ^{r^{2}-n^{2}}\left(
i\right)  ^{\alpha\left(  j\right)  }\frac{\sqrt{\left\vert S\mathrm{\det
}\mathbb{G}\right\vert }}{S\mathrm{\det}\mathbb{G}}\int_{\mathbb{R}^{m|n}}%
\rho_{\left(  r|j\right)  }(x,\theta,\eta,b)e^{i\left(  dxA\eta^{\prime
}+d\theta Bb^{\prime}\right)  }[d^{n}\eta^{\prime}d^{m}b^{\prime
}]\label{superhodge2}%
\end{equation}

As an example, we apply the definitions to the $(3|2)$ case. We adopt the
block diagonal supermetric represented by a block diagonal even super matrix
with the upper-left block given by $3\times3$ symmetric constant matrix $A$
and the lower-right block by a $2\times2$ antisymmetric constant matrix $B.$
We obtain (the wedge symbol is as usual omitted):
\begin{align}
\star1 &  =\sqrt{\left\vert \frac{\mathrm{det}(A)}{\mathrm{det}(B)}\right\vert
}\epsilon_{mnp}dx^{m}dx^{n}dx^{p}\delta^{2}(d\theta)\,,~~~~~ &  &
\in~~~\Omega^{(3|2)}\nonumber\label{sus3A}\\
\star dx^{m}= &  \sqrt{\left\vert \frac{\mathrm{det}B}{\mathrm{det}%
A}\right\vert }A^{mn}\epsilon_{npq}dx^{p}dx^{q}\delta^{2}(d\theta)\,,~~~~~ &
&  \in~~~\Omega^{(2|2)}\nonumber\\
\star d\theta^{\alpha}= &  \sqrt{\left\vert \frac{\mathrm{det}B}%
{\mathrm{det}A}\right\vert }B^{\alpha\beta}\epsilon_{mnp}dx^{m}dx^{n}%
dx^{p}\iota_{\beta}\delta^{2}(d\theta)~~~~~ &  &  \in~~~\Omega^{(2|2)}%
\,,\nonumber\\
\star dx^{m}dx^{n}= &  \sqrt{\left\vert \frac{\mathrm{det}B}{\mathrm{det}%
A}\right\vert }A^{mp}A^{nq}\epsilon_{pqr}dx^{r}\delta^{2}(d\theta)~~~~~ &  &
\in~~~\Omega^{(1|2)}\,,\nonumber\\
\star dx^{m}d\theta^{\alpha}= &  \sqrt{\left\vert \frac{\mathrm{det}%
B}{\mathrm{det}A}\right\vert }A^{mp}B^{\alpha\beta}\epsilon_{pqr}dx^{q}%
dx^{r}\iota_{\beta}\delta^{2}(d\theta)~~~~~ &  &  \in~~~\Omega^{(1|2)}%
\,,\nonumber\\
\star d\theta^{\alpha}d\theta^{\beta}=\sqrt{\left\vert \frac{\mathrm{det}%
B}{\mathrm{det}A}\right\vert } &  B^{\alpha\gamma}B^{\beta\delta}%
\epsilon_{pqr}dx^{p}dx^{q}dx^{r}\iota_{\gamma}\iota_{\delta}\delta^{2}%
(d\theta)~~~~~ &  &  \in~~~\Omega^{(1|2)}\,,
\end{align}
where $A^{mn}$ and $B^{\alpha\beta}$ are the components of the inverse
matrices of $A$ and $B$ introduced above.

If, in addition to supersymmetry, we
also impose Lorentz covariance, then $A^{mn}=A_{0}\eta^{mn}$ and
$B^{\alpha\beta}=B_{0}\epsilon^{\alpha\beta}$. Notice that in order to respect
the correct scaling behaviour, assuming that $\theta$ scales with half of the
dimension of $x$'s, $A_{0}$ has a additional power in scale dimensions w.r.t.
$B_{0}$. The quantities $A_{0}$ and $B_{0}$ are constant.

In the following, we will consider also a more general even super metric:
\begin{equation}
\mathbb{G}=\left(
\begin{array}
[c]{cc}%
G_{(ab)}(x,\theta) & G_{a\beta}(x,\theta)\\
G_{\alpha b}(x,\theta) & G_{[\alpha\beta]}(x,\theta)
\end{array}
\right)  \equiv\left(
\begin{array}
[c]{cc}%
A & C\\
D & B
\end{array}
\right)  \label{superME}%
\end{equation}
where $G_{(ab)}(x,\theta),G_{[\alpha\beta]}(x,\theta)$ are even matrices and
$G_{a\beta}(x,\theta),G_{\alpha b}(x,\theta)$ are odd matrices. In matrix
notation the even section $\sigma$ is in this case given by:%
\[
\sigma=dZ\mathbb{GG}^{-1}W=dxA\eta^{\prime}+d\theta Bb^{\prime}+dxCb^{\prime
}+d\theta D\eta^{\prime}%
\]

In general, the super matrix $\mathbb{G}$ can be expressed 
in terms of the supervielbein  $\mathbb{V}$ as follows
\begin{eqnarray}
\label{superVE}
\mathbb{G} = \mathbb{V} \mathbb{G}_0 \mathbb{V}^T 
\end{eqnarray}
where $\mathbb{G}_0$ is 
an invariant constant super matrix characterizing the tangent space 
of the supermanifold $\mathbb{R}^{(n|m)}$. The overall coefficient of the 
Hodge dual becomes 
\begin{eqnarray}
\label{superVEA}
\frac{\sqrt{|{\rm Sdet} \mathbb{G}|}}{ {\rm Sdet} \mathbb{G}} = \frac{\sqrt{|{\rm Sdet} \mathbb{G}_0|}}{{\rm Sdet}\mathbb{V} \, {\rm Sdet}\mathbb{G}_0} 
\end{eqnarray}
where ${\rm Sdet}\mathbb{V}$ is the superdeterminant of the supervielbein. 


\subsection{Metric and Supergroups}

The definition of the Hodge dual is based on the existence of a supermetric,
denoted in the previous sections (and in our previous paper \cite{Castellani:2015paa})
by $\mathbb G$.

In general, the entries of the super matrix $\mathbb G$ are superfields,
and in particular the off-diagonal blocks (for instance $G_{a\a}(x,\theta)$ are anticommuting superfields).
If we disregard constant anticommuting parameters (which {\it per se} 
have no physical interpretation), the off-diagonal
blocks are proportional to $\theta$'s. Looking for simple examples of supermetrics, it is convenient to
search among constant super matrices; thus, they must be block diagonal.

Notice that the mass dimension of each block of the matrix are different.
For example, assigning dimension 1 to the fundamental
differential $\Pi^a = dx^a + \frac12 \theta \gamma^a d\theta$ and dimension $1/2$ to the
differential $\psi^\a = d\theta^\a$, we deduce that for a dimensional-homogeneous expression
for the length element, we have to assign dimension $0$ to $G_{ab}$, $-1/2$ to $G_{a\b}$ and $G_{\a b}$,
and finally dimension $-1$ to $G_{\a\b}$. Then, $G_{a b} = \delta_{ab}$ (assuming a normalisation to 1)
and $G_{[\a\b]} = \Lambda \epsilon_{\a\b}$ where $\Lambda$ is a dimensionful parameter.
In that case, the matrix $\mathbb G$ becomes
\begin{equation}\label{HSbA}
\mathbb G =
\left(
\begin{array}{cc}
 \delta_{ab} & 0    \\
 0 & \Lambda \epsilon_{\a \b}
\end{array}
\right)
\end{equation}

A convenient method to construct meaningful examples of such a metric, and therefore a corresponding Hodge dual
for the underlying supermanifold, is given by considering the case of supergroup manifolds. For example,
given the generators $P^a, Q^\a$, bosonic and fermionic
respectively, and their commutation relations of the super algebra ${\rm osp}(1|2)$
\begin{eqnarray}\label{HSc}
[P^a, P^b] =  {\Lambda} \epsilon^{ab}_{~~c} P^c\,, ~~~~~~~
[P^{a}, Q^{a} ] = {\Lambda} (\gamma^{a})^\a_{~\b} Q^{\b}\,, ~~~~~
\{Q^{\a}, Q^{\b} \} = (\gamma_a)^{\a\b} P^a \,
\end{eqnarray}
the coordinates of the supermanifold ($x^a, \theta^\a)$ are represented by a group element
$g = e^{x_a P^a + \theta_\a Q^\a}$. The supervielbeins $(e_a, \psi_\a)$
of the supermanifolds are constructed by means of the Cartan-Maurer forms
\begin{equation}\label{HScA}
g^{-1} d g = e_a P^a + \psi_\a Q^\a\,. 
\end{equation}
$\Lambda$ is introduced in (\ref{HSc}) by rescaling the generators $P^a \longrightarrow \Lambda P^a$ and $Q^\a \longrightarrow \sqrt{\Lambda} Q^\a$.
The metric $\mathbb{G}$ is computed as the (invariant) Killing-Cartan bilinear form of the super algebra ${\rm osp}(1|2)$.

By considering the enveloping algebra and the invariant metric ${\mathbb G}$, we construct the Casimir invariant operator
\begin{equation}\label{HSd}
C_2 = \eta_{ab} P^a P^b + \Lambda \epsilon_{\a\b} Q^\a Q^\b
\end{equation}
and, finally, representing the generators of ${\rm ops}(1|2)$ in terms
of first-derivative differential operators, $C_2$ becomes a second order differential
operator (the Laplacian). In the next section, we derive the Beltrami-Laplace differential
from the Hodge dual construction given above, and find it to coincide with the Casimir (\ref{HSd}) represented as a differential operator.  

A last remark: given a topological trivial supermanifold, it is possible to define a supermetric and the corresponding Hodge dual.
 However, if the supermanifold is endowed with a supersymmetry (which geometrically
corresponds to non-trivial torsion), then the supermetric has to be compatible with it. Therefore,
the coordinates have a certain scaling behavior and transform under certain representations of
the isometry group. This implies several restrictions on the choice of the metric and the corresponding
Hodge dual even in the case of ordinary superspace.

For example, notice that by rescaling $\Lambda \longrightarrow 0$, the generators $P^a$ form an abelian subgroup
which commutes with all the generators $P^a, Q^\a$. The latter generates the usual Poincar\'e super algebra in 3d.
The Casimir invariant operator reduces to
the first term (which commutes with all generators) and the super metric $\mathbb{G}$ is degenerate (the Berezianian diverges).

The Hodge dual for a curved supermanifold is closely related to a non-vanishing cosmological constant and for that reason is suitable for curved string background with non-vanishing Ramond-Ramond fields such as $AdS_5 \times S^5$ and $AdS_4 \times {\mathbb CP}^3$.

Finally,  it has been observed in \cite{Dolan:1999dc} that the Laplace-Beltrami differential operator for $AdS_3 \times S^3$ can be constructed in terms of the Casimir operator of a super-Lie algebra. Following our construction, the Laplace-Beltrami differential operator is derived from Hodge theory using the supermetric $\mathbb{G}$  for the group manifold 
 $AdS_3 \times S^3$. This operator coincides with the one given in  \cite{Dolan:1999dc}. 


\section{Complex of Forms}

As already discussed in previous work, the complex of forms on a supermanifold is a double complex ordered according to
two degrees: {\it the form number} and {\it the picture number}. In the following, we discuss some of the characteristics of this
complex and we define two sets of differential operators acting on the complex. We denote as {\it horizontal differentials} those
which preserve the picture number and as {\it vertical differentials} those which change it and preserve the form number.

We recall that in the present section we always use as example the supermanifold $\mathbb{R}^{(3|2)}$. 

\subsection{Horizontal Differentials: $d, d^\dagger, \Delta$.}

We consider the spaces of superforms $\Omega^{(p|0)}$ with $p\geq0$ and the
following complex
\begin{equation}
0\overset{d}{\longrightarrow}\Omega^{(0|0)}\overset{d}{\longrightarrow}%
\Omega^{(1|0)}\overset{d}{\longrightarrow}\Omega^{(2|0)}\overset
{d}{\longrightarrow}\Omega^{(3|0)}\overset{d}{\longrightarrow}\Omega
^{(4|0)}\overset{d}{\longrightarrow}\dots\label{CSA}%
\end{equation}
The dimensions of these spaces are $0,1,(2|3),(6|6),(10|10)\dots$;
when $p\geq2$ the dimension is $(4p-2|4p-2)$. The notation $(a|b)$ means that we have $a+b$ generators; $a$ is the number of commuting generators and $b$ is the number of
anticommuting generators. We list below the complete decomposition of a
generic superform in those spaces. The complex has no upper bound since we can
always increase the form degree due to the commuting differentials $d\theta^{a}$.

A given form in one of the spaces $\Omega^{(p|0)}$ can be written
in terms of the generators $\Pi^{a}$ and $d\theta^{a}$ as follows
\begin{align}
\omega^{(0|0)}  &  =\Phi(x,\theta)\,,\nonumber\label{CSB}\\
\omega^{(1|0)}  &  =A_{a}\Pi^{a}+A_{\a}d\theta^{\a}\,,\nonumber\\
\omega^{(2|0)}  &  =B_{[ab]}\Pi^{a}\Pi^{b}+B_{a\alpha}\Pi^{a}d\theta
^{\a}+B_{(\alpha\beta)}d\theta^{\a}d\theta^{\b},\nonumber\\
\omega^{(3|0)}  &  =C_{[abc]}\Pi^{a}\Pi^{b}\Pi^{c}+C_{[ab]\alpha}\Pi^{a}%
\Pi^{b}d\theta^{a}\nonumber\\
&  +C_{a(\alpha\beta)}\Pi^{a}d\theta^{\a}d\theta^{
\beta} + C_{(\alpha\beta\gamma)}d\theta
^{\a}d\theta^{\beta}d\theta^{\gamma}\,,
\end{align}
where the components are superfields. As noted above, the generator of $\Omega^{(0|0)}$ is
$1$, the generators of $\Omega^{(1|0)}$ are $(\Pi^a, d\theta^\a)$, and so on. Each component,
being a superfield, contains $6 \times 2^2$ components which are either bosonic or
fermionic. 
For instance $\omega^{(1|0)}$ is decomposed in terms of the two generators $\Pi^a, d\theta^\a$ and
their coefficients $A_a(x, \theta)$ and $A_\a(x,\theta)$ are superfields with $3 \times 2^2$ and $2 \times 2^2$
components which account for $(10|10)$ degrees of freedom
\begin{equation}\label{CSBA}
A_a(x,\theta) = A^{(0)}_a(x) + A^{(1)}_{a \a}(x) \theta^\a + A^{(2)}_{a}(x) \frac{\theta^2}{2}\,, ~~~~~
A_\a(x,\theta) = A^{(0)}_\a(x) + A^{(1)}_{\a \b}(x) \theta^\b + A^{(2)}_{\a}(x) \frac{\theta^2}{2}\,.
 \end{equation}

 Acting with the differential $d$ we obtain 
the field strengths, for example:
\begin{align}
F^{(1|0)}  &  =d\omega^{(0|0)}=\Pi^{a}\partial_{a}\Phi+d\theta^{\a}D_{\a}%
\Phi\,,\nonumber\label{CSC}\\
F^{(2|0)}  &  =d\omega^{(1|0)}=\Pi^{a}\Pi^{b}\partial_{a}A_{b}+\Pi^{a}%
d\theta^{\a}(\partial_{a}A_{\a} - D_{\a}A _{a})+d\theta^{\a}d\theta^{\b}
(D_{\a} A_\b + A_a \gamma^a_{\a \b})%
\end{align}
and similarly for higher forms.
The differential $d$ increases the form number (as usual)
\begin{equation}\label{CSCA}
d: \Omega^{(p|0)} \longrightarrow \Omega^{(p+1|0)}
\end{equation}
and it is a derivation with respect to the wedge product
\begin{equation}\label{CSCB}
\wedge: \Omega^{(p|0)} \times \Omega^{(p'|0)} \longrightarrow
\Omega^{(p+p'|0)}
\end{equation}
Notice that the wedge product between superforms behaves in the conventional manner
without touching the picture number.

On the other hand, we have to consider the integral forms $\Omega^{(p|2)}$
with $p \leq3$. They form the following complex
\begin{equation}
\label{CSAA}\dots\overset{d}{\longrightarrow} \Omega^{(-1|2)} \overset
{d}{\longrightarrow} \Omega^{(0|2)} \overset{d}{\longrightarrow}
\Omega^{(1|2)} \overset{d}{\longrightarrow} \Omega^{(2|2)} \overset
{d}{\longrightarrow} \Omega^{(3|2)} \overset{d}{\longrightarrow} 0 \,.
\end{equation}
As already discussed in \cite{Castellani:2014goa}, integral forms can admit a
negative form degree (by considering derivatives of the Dirac delta
forms), and therefore there is no lower bound in the complex, but there is
an upper bound given by $\Omega^{(3|2)}$ which is the 1-dimensional
bundle of top forms (sections of the Berezinian bundle or canonical line bundle).

Again, the forms of this complex can be described locally as follows
\begin{align}
\label{CSAAA}\omega^{(3|2)}  &  = \widetilde\Phi\Pi^{3} \delta^{2}(d\theta)
\,,\nonumber\\
\omega^{(2|2)}  &  = \widetilde A^{a} (\epsilon_{abc} \Pi^{b} \Pi^{c})
\delta^{2}(d\theta) + \widetilde A^{\alpha}\Pi^{3} \iota_{\alpha}\delta
^{2}(d\theta)\,,\nonumber\\
\omega^{(1|2)}  &  = \widetilde B^{[ab]} (\epsilon_{abc} \Pi^{c}) \delta
^{2}(d\theta) + \widetilde B^{a\alpha} (\epsilon_{abc} \Pi^{b} \Pi^{c})
\iota_{\a} \delta^{2}(d\theta) + \widetilde B^{(\alpha\beta)} \Pi^3 \iota_{\a} \iota_{\beta}\delta^{2}(d\theta
)\,,\nonumber\\
\omega^{(0|2)}  &  = \widetilde C \delta^{2}(d\theta) + \widetilde C^{[ab]
\alpha} (\epsilon_{abc}\Pi^{c}) \iota_{\alpha}\delta^{2}(d\theta)\nonumber\\
&  + \widetilde C^{a (\alpha\beta)} (\epsilon_{abc} \Pi^{b} \Pi^{c})
\iota_{\alpha}\iota_{\beta}\delta^{2}(d\theta) + \widetilde C^{(\alpha
\beta\gamma)} \Pi^{3} \iota_{\alpha}\iota_{\beta}\iota_{\gamma}\delta
^{2}(d\theta) \,,
\end{align}
where the components are superfields and $\Pi^{3} = \epsilon_{abc} \Pi^{a}
\Pi^{b} \Pi^{c}$. As can be easily seen the dimensions of these spaces are again $(1|0), (3|2), (6|6), \dots, (4 p -2| 4 p -2)$
where $p$ denotes the maximum number of derivatives on Dirac delta's. Again, each term of the decomposition is a
superfield which contains bosonic and fermionic degrees of freedom. The matching of the dimensions with those of the $\Omega^{( 3 - p|0)}$ forms
is due to the duality between the two complexes.

The differential operator $d$ acts on the space of the integral forms as usual increasing the form number
\begin{equation}\label{CSAAB}
d: \Omega^{(p|2)} \longrightarrow \Omega^{(p+1|2)}
\end{equation}
leaving untouched the picture number. Acting on integral forms, one needs the distributional rules such
as $d\theta \delta'(d\theta) = - \delta(d\theta)$ and $d\theta \delta(d\theta) =0$ to compute their differential.

Among integral forms and between integral forms and superforms,
the wedge product is consistently defined as follows
\begin{equation}\label{CSAAC}
\wedge: \Omega^{(p|2)} \times \Omega^{(p'|2)} \longrightarrow 0\,, ~~~~~~
\wedge: \Omega^{(p|2)} \times \Omega^{(p'|0)} \longrightarrow \Omega^{(p+p'|2)}\,.
\end{equation}
The first definition is dictated by the anticommuting properties of delta forms and their derivatives. If derivatives of delta forms are present, we must take into account also that
$\delta(d\theta) \delta'(d\theta) = - d\theta \delta'(d\theta) \delta'(d\theta) =0$.
\\

The proposed Hodge dual operator $\star$ is a map
\begin{equation}
\label{CSAB}\star: \Omega^{(p|0)} \longrightarrow\Omega^{(3-p|2)}%
\end{equation}
verifying $\star^{2} = +1$ (because $n=3$ and if the signature of the metric is positive).
This implies that $\star$ has an inverse and can
be used to define the codifferential operator
\begin{equation}
\label{CSAC}d^{\dagger}= \star\, d \, \star
\end{equation}
which satisfies $(d^{\dagger})^{2} = (\star d \star)^{2} = \star d^{2} \star=
0$. It acts as follows
\begin{equation}
\label{CSAD}d^{\dagger}: \Omega^{(p|0)} \overset{\star}{\longrightarrow}
\Omega^{(3-p|2)} \overset{d}{\longrightarrow} \Omega^{(4-p|2)} \overset{\star
}{\longrightarrow} \Omega^{(p-1|0)}%
\end{equation}
Notice that, since $p$ could be greater than 3, the form degree of
$\Omega^{(3-p|2)}$ could be negative.
Comparing with a similar operator in the usual geometrical setting (pure
bosonic manifold), we have
\begin{equation}
\label{CSAE}d^{\dagger}: \Omega^{(p)} \overset{\star}{\longrightarrow}
\Omega^{(3-p)} \overset{d}{\longrightarrow} \Omega^{(4-p)} \overset{\star
}{\longrightarrow} \Omega^{(p-1)}%
\end{equation}
and since there are no differential forms for $p\geq3$, there is no need for
negative degree forms. Notice that $d^{\dagger}$ maps $\Omega^{(p)}$ into
$\Omega^{(p-1)}$ reducing the form degree. The same happens for
supermanifolds.

Let us consider the action of $d^{\dagger}$
on integral forms
\begin{equation}
\label{CSAF}d^{\dagger}: \Omega^{(p|2)} \overset{\star}{\longrightarrow}
\Omega^{(3-p|0)} \overset{d}{\longrightarrow} \Omega^{(4-p|0)} \overset{\star
}{\longrightarrow} \Omega^{(p-1|2)}%
\end{equation}
again it reduces the form degree number leaving unchanged the picture number.

It is a simple exercise to compute $d^{\dagger}\omega^{(1|2)}$ using the rules
given above. Of course the codifferential depends upon the choice of the metric
used to define the Hodge dual operator.

Finally, we can construct the Laplace-Beltrami differential $\Delta$ for superforms
and integral forms
\begin{equation}\label{CSAG}
\Delta = d d^\dagger + d^\dagger d:~~ \Omega^{(p|q)} \longrightarrow \Omega^{(p|q)}\,, ~~~~~~
q=0,2\,.
\end{equation}
Consequently, $\Delta$ does not change the picture and does not mix 
$\Omega^{(p|0)}$ with  $\Omega^{(p|2)}$. 
Note that, because of the Hodge dual operation $\star$, one still needs the two complexes to
construct $\Delta$.

Let us construct $\Delta$ for a $(0|0)$-superform $\omega^{(0|0)} = \Phi(x,\theta)$. We have
\begin{eqnarray}\label{CSAH}
\Delta \Phi &=& (d d^\dagger + d^\dagger d) \Phi = d \star d \Big( \Phi \Pi^3 \delta^2(d\theta) \Big) +
d^\dagger \Big( \Pi^a \partial_a \Phi + d\theta^\a D_\a \Phi \Big) = \nonumber \\
&=& \star d \Big[ \ \sqrt{\left|\frac{\det B}{\det A}\right|} A^{a a'} \partial_a \Phi \epsilon_{a'b'c'} \Pi^{b'} \Pi^{c'} \delta^2(d\theta) +
 \sqrt{\left|\frac{\det B}{\det A}\right|} B^{\a \a'} D_\a \Phi \Pi^{3} \iota_{\a'} \delta^2(d\theta) \Big] = \nonumber \\
 &=& \star \Big(  \partial_a \Big(  \sqrt{\left|\frac{\det B}{\det A}\right|} A^{a a'} \partial_{a'} \Phi\Big) +
 D_\a \Big(  \sqrt{\left|\frac{\det B}{\det A}\right|} B^{\a \a'} D_{\a'} \Phi\Big) \Pi^3 \delta^2(d\theta)\Big) = \nonumber \\
&=&  \sqrt{\left|\frac{\det A}{\det B}\right|}\Big(  \partial_a \Big( \sqrt{\left|\frac{\det B}{\det A}\right|} A^{a a'} \partial_{a'} \Phi\Big) +
 D_\a \Big(  \sqrt{\left|\frac{\det B}{\det A}\right|} B^{\a \a'} D_{\a'} \Phi\Big)\Big)
\end{eqnarray}
which reduces to
\begin{equation}\label{CSAI}
\Delta \Phi = \Big(A^{a a'} \partial_a \partial_{a'} \Phi + B^{\a \a'} D_\a D_{\a'} \Phi\Big)\
\end{equation}
in the case of constant $A^{aa'}$ and $B^{\a\a'}$. We remark that the two terms
in $\Delta$ are weighted with two dimensionful matrices in order to take into account
the correct dimension of derivatives. As discussed above, this differential operator
appears naturally as the differential representation of the second order Casimir invariant
operator for orthosymplectic and unitary supergroups.

\subsection{Vertical Differentials: $\mathbb{Y}, \mathbb{Z}$. }

In the present section, we discuss two differential operators relevant in the study of differential forms in $\Omega^{(p|0)}$ and
$\Omega^{(p|2)}$. They act mapping superforms into integral forms and vice-versa.

The first operator is constructed as follows. Given a constant commuting vector $v_\a$ we define
the object
\begin{equation}\label{PCOa}
Y_v = v_\a \theta^\a \delta(v_\beta d \theta^\beta)\,,
\end{equation}
 with the properties
 \begin{equation}\label{PCOb}
d Y_v = 0\,, ~~~~~Y_v \neq d H^{(-1|1)}\,, ~~~~~ \delta_v Y_{v} =  d \Big(v_\a \theta^\a \delta v_\b \theta^\b \delta'(v_\gamma d\theta^\gamma) \Big)\,,
\end{equation}
where $H^{(-1|1)}$ is a pseudoform and $\delta_v$ is a variation of $v$. Notice that $Y_v$ belongs to $\Omega^{(0|1)}$ and by choosing two linear independent vectors $v^{(\a)}$, we have
\begin{equation}\label{PCObb}
\mathbb Y = \prod_{\a=1}^2 Y_{v^{(\a)}} =\epsilon_{\a_1 \a_2}\theta^{\a_1} \theta^{\a_2} \epsilon^{\a_1 \a_2} \delta(d\theta^{\a_1}) \delta(d\theta^{\a_2}) =
\theta^2 \delta^2(d\theta)\,.
\end{equation}
which is an integral form of $\Omega^{(0|2)}$.
The resulting differential operator $\mathbb Y$ is independent of $v$'s since 
$\delta(v^{(1)}_\a d\theta^\a) \delta(v^{(2)}_\b d\theta^\b) = (\det (v^{(1)}, v^{(2)})^{-1} \delta^2(d\theta)$ and $v^{(1)}_\a \theta^\a) v^{(2)}_\b \theta^\b) = \det (v^{(1)}, v^{(2)}) \theta^2$. 
  
This operator (known as Picture Changing Operator, PCO) changes the picture number
and acts on superforms by the wedge product.

For example, given $\omega$ in $\Omega^{(p|0)}$ we have
\begin{eqnarray}\label{PCOc}
{\mathbb Y} &:& \Omega^{(p|0)} \longrightarrow \Omega^{(p|2)} \nonumber \\
&& \omega \longrightarrow \omega\wedge {\mathbb Y} \,,
\end{eqnarray}
If $d \omega =0$ then $d (\omega \wedge {\mathbb Y}) =0$ (by applying the Leibniz rule), and if $\omega \neq d \eta$
it follows that also $\omega \wedge {\mathbb Y} \neq d U$ where $U$ is an integral form of $\Omega^{(p-1|2)}$.
In \cite{Catenacci:2010cs}, it has been proved that ${\mathbb Y}$ is an element of the de Rham cohomology and
is globally defined. So, given an element of the cohomogy $ \omega \in H_d^{(p|0)}$,
the integral form $\omega \wedge {\mathbb Y}$ is an element of $H_d^{(p|2)}$.

An important remark: the operator ${\mathbb Y}$ being nilpotent $\mathbb Y^2 =0$ (because of
$\theta^\a \theta^\b \theta^\gamma =0$ and because of $\delta^3(d\theta) =0$) has a non-trivial kernel; so
the operation of raising the picture number by $\mathbb Y$ is not an isomorphism between integral and super forms,
but only on the cohomologies therein.

Let us consider again the 2-form $F^{(2|0)} = d A^{(1|0)} \in \Omega^{(2|0)}$
where $A^{(1|0)} = A_a \Pi^a + A_\a d\theta^\a \in \Omega^{(1|0)}$ is a gauge field.
Then we can map its field strength $F^{(2|0)}$ into an integral form (which eventually can be integrated on
a $(2|2)$ sub-supermanifold, see \cite{Witten:2012bg})
\begin{equation}\label{PCOd}
F^{(2|0)} \longrightarrow  F^{(2|2)} = F^{(2|0)} \wedge {\mathbb Y}
\end{equation}
and satisfies the Bianchi identity $d F^{(2|2)} =0$.
Using the definition of $F^{(2|0)} $ and using $d {\mathbb Y} =0$, we have
$$
F^{(2|2)} = d \left( A^{(1|0)} \wedge {\mathbb Y} \right)  \equiv d A^{(1|2)}\,,
$$
where ${A}^{(1|2)}=  A^{(1|0)} \wedge {\mathbb Y}$ is the gauge field at picture number 2.
Then, by performing a gauge transformation on $A^{(1|0)}$, namely $\delta A^{(1|0)} = d \lambda^{(0|0)}$, we have
$$\delta {A}^{(1|2)} = d \left(\lambda^{(0|0)} \wedge {\mathbb Y}\right) $$
and therefore $\lambda^{(0|2)} = \lambda^{(0|0)} \wedge {\mathbb Y}$ is viewed
as the gauge parameter at picture number 2.

By expanding $F^{(2|0)}$ in components, we have
\begin{equation}\label{PCOe}
F^{(2|0)} \wedge {\mathbb Y}
=
\Big(\partial_a A_b \Pi^a \Pi^b + \dots + (D_\a A_\b + \gamma^a_{\a\b} A_a) d\theta^\a d\theta^\b\Big) \wedge {\mathbb Y}
\end{equation}
$$
 =\Big( \partial_{[a} A_{b]}(x,0) \theta^2  \Big)\, \Pi^a \Pi^b \delta^2(d\theta)
$$
where $A_a(x,0)$ is the lowest component of the superfield $A_a$ appearing in the superconnection $A^{(1|0)}$.
This seems puzzling since we have ``killed" the complete superfield $(A_a(x,\theta), A_\a(x,\theta))$ dependence leaving
just the first component $A_a(x,0) = A^{(0)}_\a(x)$ as given in (\ref{CSBA}). On the other side, we have to note that
$F^{(2|2)}$ has $(3|2)$ independent superfield components $(F^a, F^\a)$ while $F^{(2|0)}$ has
$(6|6)$ superfield components $(F_{[ab]}, F_{a\a}, F_{(\a\b)})$. Analogously, $A^{(1|2)}$ has $(6|6)$ superfield components
$(A^{[ab]}, A^{a\a}, A^{(\a\b)})$, while $A^{(1|0)}$ has $(3|2)$ superfield components $(A_a, A_\a)$.

To solve this problem we have to modify the definition of picture changing operator
given in (\ref{PCOa}) with a more general construction.

We consider a set of anticommuting superfields $\Sigma^\a(x,\theta)$
such that $\Sigma^\a(x,0) = 0$.
They can be normalised as $\Sigma^\a(x,\theta) = \theta^\a + K^\a(x, \theta)$ with $K^\a \approx O(\theta^2)$. Then,we define
\begin{equation}\label{PCOf}
{\mathbb Y}^{(0|2)} = \prod_{i=1}^2 \Sigma^{\a_i} \delta(d \Sigma^{\a_i}) =
\prod_{i=1}^2 \Sigma^{\a_i} \delta\Big( (\delta^{\a_i}_\b + D_\b K^{\a_i}) d\theta^{\b} + \Pi^a \partial_a\Sigma^{\a_i}\Big)
\end{equation}
$$
= \prod_{i=1}^2 \Sigma^{\a_i}
\delta\left[ (\delta^{\a_i}_\b + D_\b K^{\a_i}) \Big( d\theta^{\b} + \Pi^a (1 + D K)^{-1~\b}_\gamma\partial_a\Sigma^\gamma \Big)\right]=
$$
$$
= \frac{1}{\det (1 + DK)}\prod_{i=1}^2 \Sigma^{\a_i}
\delta\left[  \Big( d\theta^{\a_i} + \Pi^a (1 + D K)^{-1~\a_i}_\gamma\partial_a\Sigma^\gamma  \Big)\right]
$$
where $(1 + DK)$ is a $2 \times 2$ invertible matrix.
Expanding the Dirac delta form and recalling that the bosonic dimension of the space is 3, we obtain the formula
\begin{eqnarray}\label{PCOg}
{\mathbb Y}^{(0|2)} &=& H(x, \theta) \delta^2(d\theta) + K_a^{\a}(x,\theta) \Pi^a \iota_\a \delta^2(d\theta) + \nonumber\\
&+& L^{a (\a\b)}(x,\theta) \epsilon_{abc}\Pi^b \Pi^c \iota_\a \iota_\b \delta^2(d\theta) +
M^{(\a\b\gamma)}(x,\theta) \Pi^3 \iota_\a \iota_\b \iota_\gamma \delta^2(d\theta)\,,
\end{eqnarray}
where the superfields $H, K_a^{\a}, L^{a(\a\b)}$ and $M^{(\a\b\gamma)}$ are easily computed in terms of $\Sigma^\a$ and its
derivatives. Even if it is not obvious from the final expression in (\ref{PCOg}), ${\mathbb Y}$ is closed and not exact from its definition
in (\ref{PCOf}). It belongs to $\Omega^{(0|2)}$
and it is globally defined; this can be checked by decomposing the supermanifold in patches and checking that ${\mathbb Y}$ is
an element of the \v{C}ech cohomology. This was done for example in \cite{Catenacci:2010cs} for super projective varieties.
Now, if we compute the new field strength $F^{(2|2)}$ by (\ref{PCOd}), one sees that the different pieces in
(\ref{PCOg}) from ${\mathbb Y}$ pick up different contributions from $F^{(2|0)}$. For instance, the $d\theta^\a \wedge d\theta^\b$
is soaked up from the third piece in (\ref{PCOg}) with the two derivatives acting on Dirac delta function.
With this more general definition of the PCO, all components of $F^{(2|0)}$ appear in the expression 
of $F^{(2|0)} \wedge \mathbb{Y}$.

Let us consider now another operator. Taken an odd vector field $X = X^a \partial_a + X^\a D_\a$
where the coefficients $X^a$ and $X^\a$ are fermionic and bosonic, respectively, we define the usual interior differential (contraction)
\begin{equation}\label{PCOh}
\iota_X = X^\a \iota_{\partial_a} + X^\a \iota_{D_\a}
\end{equation}
acting on $\Omega^{(p|0)}$ in the conventional way. The anticommuting properties of
$X$ imply that
\begin{equation}\label{PCOi}
\iota_X \iota_X \neq 0
\end{equation}
which means that $\iota_X$ is not nilpotent. Therefore, the Cartan calculus has to be modified. A complete discussion on this point
can be found in ref.s \cite{integ,Grassi:2004tv,Grassi:2007mm}. As for the differential $d\theta^\a$, we need to introduce a
distribution-like differential operator to act on $\delta(d\theta^\a)$ in the same way as $\iota_a$ acts on $\Pi^b$,
{\it i.e.} $\iota_a \Pi^b = \delta_a^b$ (we recall that $\iota_{D_\a} d\theta^\b = \delta_\a^\b$, $\iota_a d\theta^\b =0$ and
$\iota_{D_\a} \Pi^b =0$.).
We introduce the operator $\delta(\iota_{D_\a})$ acting as follows
\begin{equation}\label{PROa}
\delta(\iota_{D_\a}) \delta(d\theta^\b) = \delta^\b_\a\,.
\end{equation}
This operator has the property of removing the Dirac delta functions and therefore
of changing the picture by lowering it.
To map cohomological classes, we need to modify it in order to
be $d$ closed  and not exact. For that we define:
\begin{equation}\label{PROb}
Z_X = \left[d, \Theta(\iota_X)\right] = \delta(\iota_X) {\cal L}_X+ \frac{1}{2} \delta'(\iota_X) \iota_{[X,X]}\,, ~~~~~
{\cal L}_X = d \iota_X - \iota_X d\,,
 \end{equation}
 where $\Theta(x)$ is the usual Heaviside (step) function.
Again, if we pick up a commuting constant vector $v$, we can write the easiest example of
$Z_X$ by setting $X = v^\a \partial_\a$ (with $\{X, X\} =0$) and we have
\begin{equation}\label{PROc}
Z_v = \delta(\iota_{v^\a \partial_\a}) v^\a \partial_\a
\end{equation}
with the properties
\begin{equation}\label{PROcA}
d Z_v =0\,, ~~~~~~   Z_v \neq d H\,,  ~~~~{\rm with} ~~ H \in \Omega^{(-1|2)}\,, ~~~~~~~
\delta_v Z_v = d \eta\,, {\rm with} ~~\eta \in \Omega^{(-1|2)}
\end{equation}
Notice that, although $Z_v$ can be formally written as $d$-closed (see eq. (\ref{PROb})),
$\Theta(\iota_X)$ is not a Dirac delta form.
As in the case of $\mathbb{Y}$, it is convenient to define the product of two $Z$'s (defined with two linear independent $v$'s),
to get
\begin{equation}\label{PROd}
{\mathbb Z} = \prod_{i=1}^2 Z_{v^{(i)}} = \epsilon_{\a\b} \delta(\iota_\a) \delta(\iota_\b) \epsilon^{\a\b} \partial_\a \partial_\b\,.
\end{equation}
where the dependence on $v$'s drops out.
This differential operator acts as follows
\begin{eqnarray}\label{PROe}
\mathbb Z&:& \Omega^{(p|2)} \longrightarrow \Omega^{(p|0)} \nonumber \\
&& \omega^{(p|2)} \longrightarrow {\mathbb Z}  \omega^{(p|2)} \,.
\end{eqnarray}
As an example, we consider the integral form $\omega^{(3|2)} = \Phi \Pi^3 \delta^2(d\theta) =
\Phi d^3x \delta^2(d\theta)$ (the last equality is due to the fact that the dependence upon $d\theta$ in $\Pi$ is cancelled because of
the delta's).  Then we find
\begin{equation}\label{PROf}
{\mathbb Z}\Big( \Phi d^3x \delta^2(d\theta) \Big) = \Big(\epsilon^{\a\b} \partial_\a \partial_\b \Phi\Big) d^3x\,.
\end{equation}
 This example is also useful to show that ${\mathbb Z}$ maps cohomologies of $H^{(p|2)}_d$ into $H^{(p|0)}_d$,
 indeed since $\omega^{(3|2)}$ is automatically closed being a top integral form, we have
 \begin{eqnarray}\label{PROg}
d  \Big[{\mathbb Z}\Big( \Phi d^3x \delta^2(d\theta) \Big)\Big] &=& d \Big(\epsilon^{\a\b} \partial_\a \partial_\b \Phi\Big) d^3x =
(dx^a \partial_a + d\theta^\gamma \partial_\gamma) \Big(\epsilon^{\a\b} \partial_\a \partial_\b \Phi\Big) d^3x \nonumber \\ &=&
 \epsilon^{\a\b} \left( \partial_\gamma \partial_\a \partial_\b \Phi \right)d^3x  d\theta^\gamma = 0
\end{eqnarray}
The term of the differential $d$ with the 1-form $dx^a$ drops out since the right hand side of (\ref{PROf}) is already a three form proportional to
$d^3x$ and the last equality follows from the fact that $\partial_\a \partial_\b \partial_\gamma =0$ since they anticommute. This implies
that also ${\mathbb Z} \omega^{(3|2)}$ is closed. In the same way, for an exact form $\omega^{(p|2)} = d \omega^{(p-1|2)}$,
one can show that $ {\mathbb Z} \omega^{(0|2)}$ is also exact.

As above, we remark that while ${\mathbb Z}$ maps cohomologies into cohomologies,
it is not an isomorphism between integral forms and superforms since it is nilpotent ${\mathbb Z}^2=0$
(because of $\delta^3(\iota_X)=0$ and $\partial^3 =0$).

\section{Constraints, Rheonomy and Cohomology}

In the present setting, given the complexes of superforms discussed above, we
can study the cohomology of the de Rham operator $d$ and that of $d^{\dagger}%
$. The relevant aspect here is that we can formulate a physical interesting
model in two ways, either starting from superforms (as it has been done so far) 
or using integral forms. This last procedure might shed new light
on the construction of supersymmetric models.

However, we have to clarify how the cohomology is understood. It is easy to
show that the de Rham cohomology on superforms, by the Poincar\'e lemma
coincides with the usual cohomology
\begin{equation}
\label{COHa}H(d, \Omega^{(n|0)}) = \mathbb{R} \delta^{n,0}%
\end{equation}
which means that the only closed and not-exact forms are those in the space
$\Omega^{{0|0}}$. However, in the space of superforms we have the following
issue: considering the space of superforms $\Omega^{(1|0)}$, we have two
independent sets of superfields $A_{a}(x, \theta)$ and $A_{\alpha}(x, \theta)$, 
containing several components. In principle, they could be identified with
some physical degrees of freedom, but, generically, they represent reducible
representations of the Lorentz group and therefore they can be identified with
different type of particles. In order to overcome this problem, one imposes
some constraints on the field strength (gauge invariant constraints) in order
to reduce the number of independent components. For example, in the case of
$A^{(1|0)}$ (given in \ref{CSB}), its field strength is displayed in
(\ref{CSC}) and denoted by $F^{(2|0)}$. In order to reduce the number of
independent components to the physical ones, the constraint\footnote{In the
rheonomic language, this is expressed by the requirement that the component
along spinorial ``legs" must be expressed in terms of the components
along vectorial ``legs". We refer to \cite{Castellani}.}
\begin{equation}
\label{COHb}\iota_{\alpha}\iota_{\beta}F^{(2|0)} = (D_{(\alpha} A_{\beta)} +
\gamma^{a}_{\alpha\beta} A_{a}) =0\,.
\end{equation}
must be imposed. Then, by solving with respect to $A_{a}(x, \theta)$, we find
that the independent components are contained in the spinorial part of the
connection $A_{\alpha}(x,\theta)$. Condition (\ref{COHb}) is an obstruction to
the Bianchi identities
\begin{equation}
\label{COHc}d F^{(2|0)} =0
\end{equation}
that can be solved in terms of a single spinorial superfield
$W^{\alpha}(x, \theta)$, and we arrive to the final result (a.k.a. rheonomic
parametrisation)
\begin{align}
\label{SM-A}F^{(2|0)} = d A^{(1|0)}  &  = F_{ab} \Pi^{a} \Pi^{b} +
d\theta\gamma_{a} W \Pi^{a} \,,\nonumber\\
d W^{\alpha}  &  = \Pi^{a} \partial_{a} W^{\alpha}- F_{ab} (\gamma^{ab}
d\theta)^{\alpha}+ d\theta^{\alpha}D \,,\nonumber\\
d D  &  = \Pi^{a} \partial_{a} D + d\theta\gamma^{a} \partial_{a} W\,.
\end{align}
with the relations (obtained by Bianchi's identities)
\begin{align}
\label{SM-B} &  \partial_{[a} F_{bc]} =0 \,,\nonumber\\
&  \partial_{\alpha}F_{ab} + (\gamma_{[a} \partial_{b]} W)_{\alpha
}=0\,,\nonumber\\
&  F_{ab} + \frac12 (\gamma_{ab})^{\alpha}_{~\beta} D_{\alpha}W^{\beta}=0
\,,\nonumber\\
&  D_{\alpha}W^{\alpha}=0\,.
\end{align}
The first scalar component of the superfield $D$ is the usual auxiliary field
for the off-shell super gauge fields. Therefore, the constraints (\ref{COHb})
trasform the Bianchi identities into non-trivial equations identifying a
single field strength $W^{\alpha}$ and the superfields $A_{\alpha}$ as the
non-trivial ingredients.

We have to mention the detailed discussion of the cohomology of
superforms (based on the seminal works \cite{Gates:1980ay,Gates:1983nr})
provided in \cite{Linch:2014iza,Randall:2014gza,Gates:2014cqa}. There it is
clarified what cohomology means in the case of superforms and a
systematic technique to compute it is provided. This amounts to fix some of
the components of the field strengths to zero and to solve the Bianchi's
identities. The cohomology is identified as a \textit{relative} cohomology
which is not trivial because of the additional constraints.

Let us now move to integral forms.\footnote{In string theory, the
vertex operators needed to construct physical amplitudes are in the BRST
cohomology and they are characterised by two quantum numbers: the
\textit{ghost} number and the \textit{picture number}. For different ghost
number, there are different cohomologies, but at a different picture number (and the same ghost number),
there are the same cohomology classes. To be more precise we can choose a
representative of the same cohomology class in any picture number. The concept
of infinite dimensional complexes of superforms is easily seen in terms of
commuting super ghost fields $\gamma,\beta$.} We have to notice that acting
with the differential $d$, we move from $\Omega^{(p|2)}$ to $\Omega^{(p+1|2)}$
increasing the form number and leaving the picture number unchanged. However,
since the number of independent generators of the spaces $\Omega^{(p|2)}$
decreases as the form number increases, we see that the condition we
get by imposing the vanishing of some field strength components are not enough
to reduce to irreducible representations (usually this consists in finding a single supermultiplet
described by a superform). Therefore, in order to reproduce the relative cohomology of
the complex of superforms, we need to use the new differential $d^{\dagger}$.
\begin{equation}
d^{\dagger} A^{(2|2)}=\widetilde{F}^{(1|2)}\,, \label{SM-C}%
\end{equation}
on which we can finally put the constraints. Notice that the dimension of
$\Omega^{(2|2)}$ is equal to the dimension of $\Omega^{(1|0)}$, and via the Hodge
dual we have a simple mapping of its components
\begin{align}
\star A^{(2|2)}  &  =\widetilde{A}^{a}\star\big(\epsilon_{abc}\Pi^{b}%
\Pi^{c})\delta^{2}(d\theta)\big)+\widetilde{A}^{\alpha}\star\big(\Pi^{3}%
\iota_{\alpha}\delta^{2}(d\theta)\big)\nonumber\label{SM-D}\\
&  =\big(\widetilde{A}^{a} \widetilde{G}_{ab}+\widetilde{A}^{\alpha}\widetilde{G}_{\alpha b}%
\big)\Pi^{b}+\big(\widetilde{A}^{a}\widetilde{G}_{a \b} +\widetilde{A}^{\alpha}\widetilde{G}_{\alpha \b}
\big)d\theta^{\b}\,. %
\end{align}
where we have collected all coefficients of the Hodge dual operation as 
follows 
\begin{eqnarray}
\label{SM-DA}
\star\big(\epsilon_{abc}\Pi^{b} 
\Pi^{c})\delta^{2}(d\theta)\big) = \widetilde{G}_{ab} \Pi^b + \widetilde{G}_{a\beta} d\theta^\beta\,, 
~~~~~~~
\star\big(\Pi^{3}\iota_{\alpha}\delta^{2}(d\theta)\big) = \widetilde{G}_{\a b} \Pi^b + \widetilde{G}_{\alpha\beta} d\theta^\beta\,. 
\end{eqnarray}
At this point we can apply the differential
operator $d$ to (\ref{SM-DA}) and finally convert it into an integral form
$\Omega^{(1|2)}$ by applying again the Hodge dual. 
The components of $d\star F^{(2|0)}$ are given by
\begin{align}
\Big(d\star F^{(2|0)}\Big)_{[ab]}  &  =
\partial_{\lbrack a}\big(\widetilde {A}^{c}\widetilde{G}_{|c|b]}+\widetilde{A}^{\gamma}\widetilde{G}_{\gamma b]}\big)\,,\nonumber\label{SM-E}\\
\Big(d\star F^{(2|0)}\Big)_{\alpha b}  &  =\partial_{a}\big(\widetilde {A}^{c}\widetilde{G}_{c\a} + \widetilde{A}^{\gamma}\widetilde{G}_{\gamma \a}\big)-D_\a
\big(\widetilde{A}^{c}\widetilde{G}_{c a}+\widetilde{A}^{\gamma}\widetilde{G}_{\gamma a}\big)\,,\nonumber\\
\Big(d\star F^{(2|0)}\Big)_{(\alpha\b)}  &  =D_{(\alpha}\big(\widetilde
{A}^{c}\widetilde{G}_{c\b)}+\widetilde{A}^{\gamma}\widetilde{G}_{|\gamma|\b)}\big)+\gamma^a_{\alpha\b}
\big(\widetilde{A}^{d}\widetilde{G}_{d a}+\widetilde{A}^{\gamma}\widetilde{G}_{\gamma a}\big)\,,
\end{align}
Therefore, the constraints needed to select the
independent components of the superfield are given by
\begin{equation}
\Big(d\star F^{(2|0)}\Big)_{(\alpha\b)}=0\,. \label{SM-F}%
\end{equation}


\subsection{Non-abelian Gauge Fields}

Having discussed the constraints to define physical degrees of freedom in terms of a given superform or integral form,
we here discuss how the non-abelian terms are constructed. In the case of superforms (which we recall are $0$-picture forms)
the construction is conventional. Given the superform $A^{(1|0)}$, we consider it with value in a given Lie algebra and
we construct its field strength as follows
\begin{equation}\label{NONA}
F^{(2|0)} = d A^{(1|0)} + \frac12 [A^{(1|0)}, A^{(1|0)}]\,,
\end{equation}
where the commutator is taken on the Lie algebra and the two forms
are multiplied with the wedge product. $F^{(2|0)}$ satisfies the Bianchi's identities
\begin{equation}\label{NONB}
\nabla_A F^{(2|0)} = 0\,.
\end{equation}
where $\nabla_A$ is the covariant derivative with respect to $A^{(1|0)}$. Notice that
multiplying two $0$-picture forms, we do not change the global picture.
The situation is rather different for integral forms. Let us consider the $2$-picture integral forms $A^{(1|2)}$ with value
in a Lie algebra. We cannot construct the field strength in the usual way since
we cannot multiply $A^{(1|2)}$ by itself (as discussed above). For that we need the PCO ${\mathbb Z}$
to reduce the picture first, namely we consider the $0$-picture connection ${\mathbb Z} A^{(1|2)}$
and one possible definition is
\begin{equation}\label{NONC}
F^{(2|2)} = d A^{(1|2)} + A^{(1|2)} \wedge {\mathbb Z} A^{(1|2)}
\end{equation}
However, by computing its Bianchi identity, we
run into some problems. Indeed, we find
\begin{eqnarray}
\label{NOND}
\nonumber
d F^{(2|2)} = d A^{(1|2)} \wedge  {\mathbb Z} A^{(1|2)}  - A^{(1|2)} \wedge  d {\mathbb Z} A^{(1|2)} =
d A^{(1|2)} \wedge  {\mathbb Z} A^{(1|2)}  - A^{(1|2)} \wedge  {\mathbb Z} d A^{(1|2)}
\end{eqnarray}
Using the definition (\ref{NONC}), we get
\begin{eqnarray}
\label{NONE}
\nonumber
d F^{(2|2)} &=& \left( F^{(2|2)} - A^{(1|2)} \wedge {\mathbb Z} A^{(1|2)}  \right) \wedge  {\mathbb Z} A^{(1|2)}  -
A^{(1|2)} \wedge  {\mathbb Z} \left( F^{(2|2)} - A^{(1|2)} \wedge {\mathbb Z} A^{(1|2)}  \right) \nonumber \\
&=& F^{(2|2)} \wedge  {\mathbb Z} A^{(1|2)}  - A^{(1|2)} \wedge  {\mathbb Z} F^{(2|2)} \nonumber \\
&-&
 A^{(1|2)} \wedge \left({\mathbb Z} A^{(1|2)}  \wedge  {\mathbb Z} A^{(1|2)}
-  {\mathbb Z} \Big(A^{(1|2)} \wedge {\mathbb Z} A^{(1|2)}\Big) \right)
\end{eqnarray}
that we can rewrite as
\begin{eqnarray}
\label{NONF}
\nonumber
\nabla_{{\mathbb Z} A^{(1|2)}} F^{(2|2)} &=& {\mathbb Z} A^{(1|2)}  \wedge F^{(2|2)} -  A^{(1|2)}  \wedge {\mathbb Z}  F^{(2|2)} \nonumber \\
&-&
 A^{(1|2)} \wedge \left({\mathbb Z} A^{(1|2)}  \wedge  {\mathbb Z} A^{(1|2)}
-  {\mathbb Z} \Big(A^{(1|2)} \wedge {\mathbb Z} A^{(1|2)}\Big) \right)
\end{eqnarray}
We notice that the first two terms on the right hand side do not cancel since
\begin{eqnarray}
\label{NONG}
\nonumber
 {\mathbb Z} A^{(1|2)}  \wedge F^{(2|2)} -  A^{(1|2)}  \wedge {\mathbb Z}  F^{(2|2)} \neq 0
\end{eqnarray}
which means that $\mathbb{Z}$ is not a derivation of the exterior algebra. In addition,
notice that $A^{(1|2)}  \wedge F^{(2|2)}  =0$ by the rule we have established before.
For the same reason, we have
\begin{eqnarray}
\label{NONH}
\nonumber
{\mathbb Z} A^{(1|2)}  \wedge  {\mathbb Z} A^{(1|2)}
-  {\mathbb Z} \Big(A^{(1|2)} \wedge {\mathbb Z} A^{(1|2)}\Big) \neq 0\,.
\end{eqnarray}
A way to solve this problem is to consider the sum of all possible pictures for a
$1$-form, namely
${\cal A}= A^{(1|0)} + A^{(1|1)}+ A^{(1|2)}$
where $ A^{(1|1)}$ is a pseudo form with a single Dirac delta function.
We have not discussed such an object in the present paper and in the past
papers, since they require some additional studies. In particular, it can be shown that for
a given form number the space of pseudo forms is infinite dimensional.
In order to construct $A^{(1|1)}$ we can act on $A^{(1|2)}$ with a single PCO $Z_v$
(introduced in (\ref{PROc}))  while the relation between $A^{(1|0)}$ and $A^{(1|2)}$
is obtained by acting with $\mathbb{Z}$ which removes both the Dirac delta functions. We construct
the field strength as usual
\begin{eqnarray}
\label{NONI}
\nonumber
{\cal F} &=& d {\cal A} + {\cal A} \wedge {\cal A}  = \left( d A^{(1|0)} + A^{(1|0)} \wedge A^{(1|0)} \right) \nonumber \\
&+& \left( d A^{(1|1)} + A^{(1|1)} \wedge A^{(1|0)} + A^{(1|0)} \wedge A^{(1|1)}  \right)  \nonumber \\
&+& \left( d A^{(1|2)} + A^{(1|2)} \wedge A^{(1|0)} + A^{(1|1)} \wedge A^{(1|1)}  + A^{(1|0)} \wedge A^{(1|2)}\right)\,,
\end{eqnarray}
which satisfies by definition to the Bianchi identities
\begin{eqnarray}
\label{NONJ}
\nonumber
&& d F^{(2|0)} + F^{(2|0)} \wedge A^{(1|0)} + A^{(1|0)} \wedge F^{(2|0)} =0 \,,\nonumber \\
&& d F^{(2|1)} + F^{(2|1)} \wedge A^{(1|0)} + A^{(1|0)} \wedge F^{(2|1)}  + F^{(2|0)} \wedge A^{(1|1)} + A^{(1|1)} \wedge F^{(2|0)} =0 \,, \nonumber \\
&& d F^{(2|2)} + F^{(2|2)} \wedge A^{(1|0)} + F^{(2|1)} \wedge A^{(1|1)} + F^{(2|0)} \wedge A^{(1|2)} \nonumber \\
&& ~~~~~~~~\, + A^{(1|0)} \wedge F^{(2|2)}  + A^{(1|1)} \wedge F^{(2|1)} + A^{(1|0)} \wedge F^{(2|2)} =0 \,.
\end{eqnarray}
The first equation involves only the contribution at zero picture and it is the conventional expression for the superforms.
The other lines invoke the presence of the gauge connection and of the field strength at different pictures.  Notice that
for consistency one needs also the pseudo-forms at picture number equal to one. We conclude that a complete theory of
gauge connections on supermanifolds and non-abelian gauge group is still missing and requires new methods for
dealing with the infinite dimensional spaces of pseudo-forms. 
This will be studied in a forthcoming publication.


\section{Dualities}

Going back to abelian connections, we discuss dualities on supermanifolds.
Let us recall what happens on a bosonic manifold ({\it i.e.} $\mathbb{R}^3$). We have the complex of differential
forms
\begin{equation}\label{duaA}
0 \overset{d}{\longrightarrow} \Omega^{(0)} \overset
{d}{\longrightarrow} \Omega^{(1)} \overset{d}{\longrightarrow}
\Omega^{(2)} \overset{d}{\longrightarrow} \Omega^{(3)} \overset
{d}{\longrightarrow} 0 \,.
\end{equation}
Then, if we start with a $0$-form $A^{(0)}$, we have the following sequence
\begin{equation}\label{duaB}
A^{(0)} \overset{d}{\longrightarrow} F^{(1)} = d A^{(0)} \overset{\star}{\longrightarrow} F^{(2)} = d A^{(1)}
\end{equation}
We start with a $0$-form and we construct its field strength $F^{(1)} = d A^{(0)}$, which
obviously satisfies the Bianchi identity $ d F^{(1)} =0$. Its field strength is a $1$-form. Now, we consider its Hodge dual, which in
three dimensions corresponds to a $2$-form $F^{(2)}$. If we require that the $F^{(2)}$ is closed, 
the Poincar\'e lemma implies
that $F^{(2)} = d A^{(1)}$. However, the closure of $F^{(2)}$ implies the co-closure of $F^{(1)}$, namely $d \star F^{(1)} =0$.
This, combined with the Bianchi identities, implies $d \star d A^{(0)} =0$ which is the Klein-Gordon equation in three dimensions.  On the other
side, we have that the Bianchi identities for $F^{(1)}$ imply also that $d \star F^{(2)} =0$, which, combined with its Bianchi identities, leads to
$d\star d A^{(1)} =0$ which is the Maxwell equation for a vector field. Together with the gauge invariance of $A^{(1)}$, namely
$\delta A^{(1)} = d\lambda^{(0)}$, we have that $A^{(1)}$ describes a single propagating degree of freedom for a massless gauge boson in three dimensions.
This is  the basic argument for the duality between a scalar field in three dimensions and a massless gauge boson.

In the same way, one can easily prove that in three dimensions there are no higher $p$-forms with propagating degrees of freedom. Indeed, if
one starts from a $2$-form $A^{(2)}$, its field strength is a $3$-form $F^{(3)}$ and its Hodge dual 
is a $0$-form, implying that there is no propagating degree of freedom. Notice that this argument 
works for on-shell fields. This is essential to guarantee the correct
matching of degrees of freedom in each multiplet.

In the case of supermanifolds, there is the problem of the unboundness of the complex of superforms. Therefore, it is not clear whether additional propagating
supermultiplets can be described by higher rank $p$-superforms. On the other side, we know that the physical content of the
complex of superforms should be mirrored into the complex of integral forms.

If we start from a $(0|0)$-form $A^{(0|0)}$, which describes a scalar superfield with $(2|2)$ degrees of freedom (as can be checked easily by counting
the independent coefficients of the $\theta$ expansion), we can build its field strength as above
\begin{equation}\label{duaC}
A^{(0|0)} \overset{d}{\longrightarrow} F^{(1|0)} = d A^{(0|0)} \overset{\star}{\longrightarrow} F^{(2|2)} = d A^{(1|2)}
\end{equation}
The resulting superform $F^{(1|0)} = d A^{(0|0)}$ satisfies the Bianchi identities. Consequently, the Hodge dual $F^{(2|2)}$
is an integral form and its closure implies its exactness by the Poincar\'e lemma which is valid
in the case of $\Omega^{(p|2)}$ with $p>0$ . Then, by repeating
the same argument as above, we have that the Bianchi identity on $F^{(2|2)}$ implies the equation
\begin{equation}\label{duaD}
d \star d  A^{(0|0)} =0\,,
\end{equation}
namely, it leads to a modified Klein-Gordon equation (with high derivative terms (see \cite{Castellani:2015paa}). It describes
a scalar superfield with high derivative terms.
On the other hand, we have that the Bianchi identity on $F^{(1|0)}$ implies the
following equation
\begin{equation}\label{duaDA}
d \star d  A^{(1|2)} =0\,.
\end{equation}
One finds that it describes an on-shell gauge supermultiplet
 (made by one bosonic degree of freedom and one fermionic degree of freedom). 
 The on-shell condition, the
 supersymmetry and the gauge symmetry guarantee that indeed we are describing such a supermultiplet. We have to
 notice that this is not the usual representation of a gauge supermultiplet (which is normally described by a $(1|0)$
 superform $A^{(1|0)}$) and that it corresponds to the mirror representation in terms of integral forms of the same multiplets contained in the complex of superforms.

 On the other hand, if we start from the conventional superform $A^{(1|0)}$ for a gauge multiplet, we
 have the following sequence
\begin{equation}\label{duaCA}
A^{(1|0)} \overset{d}{\longrightarrow} F^{(2|0)} = d A^{(1|0)} \overset{\star}{\longrightarrow} F^{(1|2)} = dA^{(0|2)}
\end{equation}
where $A^{(0|2)}$ is a $(0|2)$ integral form. Again, by exploiting the Bianchi identities on both sides of the Hodge duality, we
find the equivalence between the two descriptions.

Finally, let us study what happens if we further increase the form number (which is indeed possible without any restriction
in the case of superforms). We have
\begin{equation}\label{duaDB}
A^{(p|0)} \overset{d}{\longrightarrow} F^{(p+1|0)} = d A^{(p|0)} \overset{\star}{\longrightarrow} F^{(2-p|2)} =d A^{(1-p|2)}
\end{equation}
If $p \geq 2$, we see that $A^{(1-p|2)}$ belongs to the space of integral forms with negative form degree. However, this  space cannot contain any
physical degree of freedom since, by means of the PCO operators (which map physical degrees of freedom into physical degrees of freedom)
we see  that there is no room for any physical content in that space. Thus, we conclude that higher-rank $A^{(p|0)}$ superforms cannot
describe on-shell degrees of freedom consistently in the bosonic three dimensional case.

This construction resolves the issues of higher rank superforms and the description of dual superfields.

\section*{Summary}
We have generalized the construction of the Hodge dual via Fourier transform given in a 
previous work. We apply the Hodge dual theory to construct the Laplace-Beltrami differential and 
to analise the Hodge dualities  in supersymmetric models. We discuss the relation between 
superspace constraints for superforms and for integral forms. In addition, we discuss the relation  
between integral forms and superforms via PCO's and, in that framework, we consider non-abelian gauge fields in the space of integral forms.  



\begin{thebibliography}{99}                                                                                               %


\bibitem {Voronov2}T.~Voronov and A.~Zorich: {``Integral transformations of
pseudodifferential forms''}, Usp. Mat. Nauk, 41 (1986) 167-168; T.~Voronov and
A.~Zorich: {``Complex of forms on a supermanifold''}, Funktsional. Anal. i
Prilozhen., 20 (1986) 58-65; T.~Voronov and A.~Zorich: {``Theory of bordisms and
homotopy properties of supermanifolds''}, Funktsional. Anal. i Prilozhen., 21
(1987) 77-78; T.~Voronov and A.~Zorich: {``Cohomology of supermanifolds and
integral geometry''}, Soviet Math. Dokl., 37 (1988) 96-101.

\bibitem {integ}A.~Belopolsky, ``New geometrical approach to superstrings'',
[arXiv:hep-th/9703183]. \ A.~Belopolsky, ``Picture changing operators in
supergeometry and superstring theory'', [arXiv:hep-th/9706033].


\bibitem {Berkovits:2004px}N.~Berkovits, ``Multiloop amplitudes and vanishing
theorems using the pure spinor formalism for the superstring,'' JHEP
\textbf{0409} (2004) 047 [hep-th/0406055].

\bibitem{Berkovits:2009gi}
  N.~Berkovits and W.~Siegel,
  ``Regularizing Cubic Open Neveu-Schwarz String Field Theory,''
  JHEP {\bf 0911} (2009) 021
  [arXiv:0901.3386 [hep-th]].

\bibitem {Catenacci}R.~Catenacci, M.~Debernardi, P.~A.~Grassi and D.~Matessi,
``Balanced Superprojective Varieties'', J. Geo. and Phys. Volume 59, Issue 10
(2009) p. 1363-1378. [arXiv:0707.4246 [hep-th]].


\bibitem {Catenacci:2010cs}R.~Catenacci, M.~Debernardi, P.~A.~Grassi and
D.~Matessi, ``\v{C}ech and de Rham Cohomology of Integral Forms,''
J.\ Geom.\ Phys.\ \textbf{62} (2012) 890 [arXiv:1003.2506 [math-ph]].


\bibitem {Witten:2012bg}E.~Witten, ``Notes On Supermanifolds and
Integration,'' arXiv:1209.2199 [hep-th].

\bibitem {Castellani:2014goa}L.~Castellani, R.~Catenacci and P.~A.~Grassi,
``Supergravity Actions with Integral Forms,'' Nucl.\ Phys.\ B \textbf{889},
419 (2014) [arXiv:1409.0192 [hep-th]].

\bibitem {Castellani:2015paa}L.~Castellani, R.~Catenacci and P.~A.~Grassi,
``The Geometry of Supermanifolds and New Supersymmetric Actions,''
arXiv:1503.07886 [hep-th].

\bibitem{Dolan:1999dc} 
  L.~Dolan and E.~Witten,
  ``Vertex operators for AdS(3) background with Ramond-Ramond flux,''
  JHEP {\bf 9911}, 003 (1999)
  [hep-th/9910205].


\bibitem {Gates:1983nr}S.~J.~Gates, M.~T.~Grisaru, M.~Rocek and W.~Siegel,
``Superspace Or One Thousand and One Lessons in Supersymmetry,''
Front.\ Phys.\ \textbf{58}, 1 (1983) [hep-th/0108200].


\bibitem{Grassi:2004tv}
  P.~A.~Grassi and G.~Policastro,
  ``Super-Chern-Simons theory as superstring theory,''
  hep-th/0412272.


\bibitem{Grassi:2007mm}
  P.~A.~Grassi and M.~Marescotti,
  ``Integration of Superforms and Super-Thom Class,''
  arXiv:0712.2600 [hep-th].


\bibitem {Castellani}L.~Castellani, R.~D'Auria and P.~Fr\'e, ``Supergravity
and superstrings: A Geometric perspective,'' in 3 vol.s, Singapore: World
Scientific (1991) 1375-2162;


\bibitem {Gates:1980ay}S.~J.~Gates, Jr., ``Super P Form Gauge Superfields,''
Nucl.\ Phys.\ B \textbf{184}, 381 (1981).



\bibitem {Linch:2014iza}W.~D.~Linch and S.~Randall, ``Superspace de Rham
Complex and Relative Cohomology,'' arXiv:1412.4686 [hep-th].


\bibitem {Randall:2014gza}S.~Randall, ``The Structure of Superforms,''
arXiv:1412.4448 [hep-th].


\bibitem {Gates:2014cqa}S.~J.~Gates, W.~D.~Linch and S.~Randall, ``Superforms
in Five-Dimensional, $N = 1$ Superspace,'' arXiv:1412.4086 [hep-th].




\end{thebibliography}
\end{document}